\documentclass[12pt]{iopart}
\usepackage{iopams}  

\usepackage{graphicx,upgreek,amsfonts,textcomp, wasysym}

\usepackage[french,english]{babel}
\usepackage[utf8x]{inputenc}
\usepackage[T1]{fontenc}
\usepackage[usenames]{color}

\usepackage[usenames,dvipsnames]{xcolor}

\definecolor{darkblue}{rgb}{0,0,0.6}
\definecolor{darkred}{rgb}{0.6,0,0}

\usepackage{hyperref}
\hypersetup{
bookmarksopen=true,
pdftitle="",
pdfauthor="", 
pdfsubject="", 
pdfstartview={FitH},		
pdfmenubar=true,			
pdfhighlight=/O,			
colorlinks=true,			
urlcolor=black,
citecolor=black,
linkcolor=black,
}
\newcommand{\figref}[1]{Fig.~\ref{#1}}
\newcommand{\eqref}[1]{Eq.~(\ref{#1})}

\begin{document}

\title[Statistics of roughness parameters]{A numerical study of the statistics of roughness parameters for fluctuating interfaces
}

\author{S. Bustingorry}
\address{Instituto de Nanociencia y Nanotecnología, CNEA--CONICET, Centro At\'omico Bariloche, Av. Bustillo 9500, R8402AGP San Carlos de Bariloche, R\'{\i}o Negro, Argentina}
\ead{sbusting@cab.cnea.gov.ar}
\author{J. Guyonnet}
\address{DQMP, University of Geneva, 24 Quai Ernest Ansermet, CH--1211 Geneva 4, Switzerland}
\author{P. Paruch}
\address{DQMP, University of Geneva, 24 Quai Ernest Ansermet, CH--1211 Geneva 4, Switzerland}
\author{E. Agoritsas}
\address{Institute of Physics, Ecole Polytechnique F{\'e}d{\'e}rale de Lausanne (EPFL), CH--1015 Lausanne, Switzerland}

\date{\today}
\vspace{10pt}
\begin{indented}
\item[] \today
\end{indented}


\begin{abstract}

Self-affine rough interfaces are ubiquitous in experimental systems, and display characteristic scaling properties as a signature of the nature of disorder in their supporting medium, \textit{i.e.}~of the statistical features of its heterogeneities.
Different methods have been used to extract roughness information from such self-affine structures, and in particular their scaling exponents and associated prefactors.
Notably, for an experimental characterization of roughness features, it is of paramount importance to properly assess sample-to-sample fluctuations of roughness parameters.
Here, by performing scaling analysis based on displacement correlation functions in real and reciprocal space, we compute statistical properties of the roughness parameters.
As an ideal, artifact-free reference case study
and particularly targeting finite-size systems,
we consider three cases of numerically simulated one-dimensional interfaces:
\textit{(i)}~elastic lines under thermal fluctuations and free of disorder,
\textit{(ii)}~directed polymers in equilibrium with a disordered energy landscape,
and \textit{(iii)}~elastic lines in the critical depinning state when the external applied driving force equals the depinning force set by disorder.
Our results shows that sample-to-sample fluctuations are rather large when measuring the roughness exponent. These fluctuations are also relevant for roughness amplitudes. Therefore a minimum of independent interface realizations (at least a few tens in our numerical simulations) should be used to guarantee sufficient statistical averaging, an issue often overlooked in experimental reports.

\end{abstract}

%
\vspace{2pc}

%
%
%
%

\section{Introduction}
\label{sec_intro}

Ferroic materials are characterized by a spontaneous order parameter that can be reversibly switched between at least two energetically-equivalent ground states by an appropriate conjugated field. For example, in ferroelectrics and ferromagnets these order parameters are the polarization and the magnetization, respectively, switchable by applying an electric or magnetic field. Regions of homogeneous order parameter state in the sample are called domains, separated by nanoscale boundaries known as domain walls. The ability to controllably engineer ferroic domains in increasingly miniaturized devices has played a significant role in the integration of these materials into the electronics industry~\cite{polla_97_mems,kumar_apl_04_SAW, scott_sci_89_memories,waser_natmat_04_memories, Parkin2008, Hayashi2008, Allwood2005}. At the most fundamental level, such engineering is built on the understanding and control of the static and dynamical behavior of the domain walls, which determine the switching, growth, stability, and shape of ferroic domains \cite{lemerle_prl_98_FMDW_creep, paruch_jap_06_dynamics_FE, paruch_dw_review_07, metaxas_depinning_thermal_rounding, Metaxas2010}. 

One extremely useful theoretical approach to study domain walls in ferroic materials is to model them as fluctuating elastic manifolds subject to the spatial inhomogeneities of an underlying disordered potential~\cite{giamarchi_domainwall_review}.
A remarkable feature of this reductionist picture is that, because the underlying microscopic details of the system are only considered through a few effective parameters, it can be applied to systems as diverse as surface growth phenomena~\cite{barabasi_surface_growth_95}, fracture surfaces~\cite{mandelbrot_nature_84_cracks_metal}, burning~\cite{myllys_prl_00_burning_fronts} and wetting~\cite{rubio_prl_89_wetting_fronts} fronts, edges of bacterial colonies~\cite{bonachela_jstatphys_11_bacterias_DES}, cell migration \cite{Chepizhko_pnas_16_cell_front}, cell membranes~\cite{speck_pre_12_cell_membrane}, as well as ferroic domain walls~\cite{lemerle_prl_98_FMDW_creep,paruch_13_FE_DW_review,Ferre_cras_13_DW}.
In this approach, the complex static and dynamical properties of the interface emerge from a seemingly simple competition between elasticity, temperature and disorder pinning.  In particular, such \emph{disordered elastic systems} present a rough morphology with characteristic self-affine scaling properties, which depend on the dimensionality of the system, the range of the elastic interactions, and the nature of the disorder~\cite{agoritsas_physb_12_DES}.
The quantitative characterization of this roughness, including the value of the associated scaling roughness exponent $\zeta$, can rely on several methods either in real or in reciprocal space.

Experimental roughness studies in ferroic materials~\cite{lemerle_prl_98_FMDW_creep, paruch_prl_05_dw_roughness_FE, Bauer2005, catalan_prl_08_BFO_DW, pertsev_jap_11_ceramics, xiao_apl_13_PVDF_roughness_creep, Moon2013, Domenichini2019, DiazPardo2019, Jordan2020} have generally used real-space analysis of such domain walls, built on images covering a finite number of pixels, typically of the order of a few hundreds, and thus always requiring a detailed assessment of finite-size effects.
More importantly, real-space methods are mainly used to extract the value of the roughness exponent $\zeta$ from the power-law growth of the correlation function of relative displacements.
As shown in a comparative study of analysis methods on numerical --~and thus exactly defined~-- self-affine profiles, the accuracy of $\zeta$ estimation can in fact vary significantly depending on the method used~\cite{schmittbuhl_pre_95_reliability_roughness}.
Furthermore, adequate statistical averaging is an absolutely critical issue, with trustworthy $\zeta$ estimates obtained only when considering at least a few tens of independent mono-affine interfaces~\cite{Jordan2020, guyonnet_prl_12_multiscaling}. 
In order to assess roughness features as the roughness exponent $\zeta$, particularly in experimental situations where the number of interfaces is finite, a thorough evaluation of statistical fluctuations should be considered. This is specially important when comparing roughness exponent values, obtained for different materials and in different experimental conditions.
Such an evaluation would also allow the establishment of a well-defined analysis protocol, which could be applied over all the different ferroic systems under investigation and moreover to the general class of interfaces described as disordered elastic systems~\cite{giamarchi_domainwall_review, barabasi_surface_growth_95, agoritsas_physb_12_DES, Brazovskii2004}.

We perform here an evaluation of statistical fluctuations using three different numerical models of fluctuating interfaces, well benchmarked from previous works, corresponding to different universality classes.
We first found that the mean value of individual roughness parameters, \textit{i.e.}~each from a single independent interface, converges to the roughness parameters obtained using averaged correlation functions. Notably, the distributions of roughness parameters from independent interfaces, an experimentally relevant measure, are appreciably wide and size-dependent. Finally, our numerical results show that a set with at least a few tens of independent interface realizations should be used to obtain representative averaged values for the roughness exponent, something usually overlooked in experimental reports.

The rest of the manuscript is organized as follows.
In Sec.~\ref{sec_roughness_analysis} we introduce the main definitions used as a metrics for the characterization of interfaces fluctuations and discuss some key features.
Then, Sec.~\ref{sec_numerical} introduces the three studied models and present the main numerical results, first showing roughness parameters obtained using averaged correlation functions and then presenting the statistics of single interface roughness parameters.
Finally, a brief discussion and a summary of the results is presented in Sec.~\ref{sec_concl}.

\section{Roughness metrics and scaling analysis}
\label{sec_roughness_analysis}

Since the seminal work of Mandelbrot \textit{et al.} revealing the self-affine nature of cracks in metals~\cite{mandelbrot_nature_84_cracks_metal}, a significant number of different methods have been established and used to quantify the roughness of self-affine interfaces, focusing in particular on fracture surfaces~\cite{maloy_prl_92_crack,schmittbuhl_jgeophys_95_crack_scaling}. 
The key quantity to be determined is the roughness exponent $\zeta$, which characterizes the geometrical properties of interfaces through the power-law growth of their transverse fluctuations $w$ with respect to the longitudinal size of the interface $\ell$, \textit{i.e.}~${w(\ell) \approx b \, \ell^\zeta}$. The roughness exponent is expected to characterize universal behavior~\cite{barabasi_surface_growth_95}, with well defined values associated to each universality class.
In addition to the roughness exponent, roughness information is contained within the prefactor $b$ accompanying power-law growth. This is called the roughness amplitude and is not expected to be universal, only containing information about the intrinsic disorder of the system. Both the roughness exponent and the roughness amplitude are the roughness parameters we are measuring in different model systems.

In all the proposed methods for the determination of roughness parameters, a complete knowledge of the interface position is assumed, in which case the analysis of the roughness can be carried out via either reciprocal-space or real-space autocorrelation functions.
We restrict ourselves to effective one-dimensional interfaces, as they are particularly relevant for many experimental ferroic domain walls, but the following definitions can be generalized to higher dimensions.

In this section, we briefly recall the basic definitions of the roughness fluctuations we use to measure roughness parameters and discuss some key properties.

\subsection{Measuring the roughness parameters}
\label{sec_defs}

\begin{figure}[!thbp]
\begin{center}
\includegraphics[width=0.8\columnwidth]{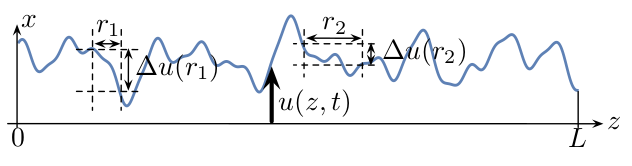}
\end{center}
\caption{
Profile of a one-dimensional interface, parameterized by the displacement field ${u(z,t)}$.  For illustation purposes the relative displacements $\Delta u(r)$ are shown for $r$ taking the values $r_1$ and $r_2$. The variance of  the relative displacements of the profile ${\left\lbrace \Delta u(r,t) \right\rbrace}$ is given by the displacement-displacement correlation function ${B(r)}$.
}
\label{fig:def-roughness}
\end{figure}

In a general sense, the roughness of an interface characterizes its geometrical fluctuations~\cite{barabasi_surface_growth_95}.
Here we specify particular real-space and reciprocal-space definitions of correlation functions giving alternative access to the roughness parameters, all relying on the displacement field ${u(z)}$ which parameterized a given configuration of an interface at time $t$ with respect to an arbitrary reference configuration, as illustrated in Fig.~\ref{fig:def-roughness}.
A usual assumption in the theoretical framework of disordered elastic systems is that the interface has no overhangs, so that ${u(z)}$ is univalued~\cite{agoritsas_physb_12_DES}.

Real-space fluctuations can be defined in terms of different correlation functions, as the global or local width.
Another local quantity containing geometrical information on interfaces is the displacement-displacement correlation function, sometimes referred to as the height-height correlation function, the height-difference correlation function, or simply the \emph{roughness function}:
\begin{equation}
 B(r)
 = \overline{\langle [u(z+r) - u(z)]^2 \rangle_L},
 \label{equ_br}
\end{equation}
where  ${\Delta u(r) =u(z+r) - u(z)}$ is the relative transverse displacement between pairs of sites a distance $r$ apart, as illustrated in Fig.~\ref{fig:def-roughness}, and ${B(r)}$ is simply the variance of the probability distribution function of relative displacements ${\mathcal{P}(\Delta u(r) )}$.
In \eqref{equ_br} $\langle \cdots \rangle_L$ stands for an average over different $z$ values for a single profile of size $L$, while $\overline{\cdots}$ corresponds to an average over different realizations.
 Notice that if we consider a set of $N$ independent realizations of interfaces profiles $u_i(z)$, with $i = 1,2,...,N$, and  $B_i(r) = \langle [u_i(z+r) - u_i(z)]^2 \rangle_L$ is the roughness function of the interface with label $i$  , then averaging over realizations means that $B(r) = \overline{B_i(r)}$.

The roughness function provides a convenient way to experimentally measure the roughness exponent, and has thus been used as a primary analysis tool in ferroic systems \cite{lemerle_prl_98_FMDW_creep, paruch_prl_05_dw_roughness_FE, catalan_prl_08_BFO_DW, Jordan2020, Jost_pre_98_roughness_B, paruch_prb_12_quench}.
It is usually assumed that the system is in a stationary situation where the time-dependence can be ignored. Denoting by $\xi(t)$ to the growing fluctuations correlation length, the stationary limit corresponds to $\xi(t)\gg r$, as expected for the classic Family-Vicsek scaling scenario~\cite{barabasi_surface_growth_95}. Then the roughness function is expected to behave as
\begin{equation}
 \label{equ_Br_scaling}
 B(r) = B_0 \left( \frac{r}{r_0} \right)^{2 \zeta},
\end{equation}
where $B_0$ and $\zeta$ are the real-space roughness amplitude and roughness exponent, respectively, \textit{i.e.}~the roughness parameters. The scale $r_0$ is a system dependent reference length scale so that $B_0$ has dimensions of length squared.

An alternative option to real-space correlation functions is to compute correlations in reciprocal space. A particularly useful quantity is the displacement power spectrum, referred to as the \emph{structure factor}:
\begin{equation}
 S(q) = \overline{\widetilde{u}(q) \widetilde{u}(-q)},
\label{equ_sq}
\end{equation}
where
\begin{equation}
 \widetilde{u}(q) = \frac{1}{L} \int_0^L dz \, u(z) \, e^{-i q z}
\end{equation}
is the Fourier transform of the displacement field $u(z)$ defining the interface position. Formally, the structure factor ${S(q)}$ and the roughness function ${B(r)}$ contain the same geometrical information and are related through
\begin{equation}
 B(r)
 = \int \frac{dq}{\pi} \, \left[1-\cos\left(qr\right)\right] \, S(q).
\label{equ_br_sq}
\end{equation}
For stationary self-affine interfaces $q \ll \xi(t)^{-1}$ and the structure factor scales as
\begin{equation}
 \label{equ_Sq_scaling}
 S(q) = S_0 \left( \frac{q}{q_0} \right)^{-(1+2 \zeta)},
\end{equation}
where $S_0$ is the reciprocal-space roughness amplitude and $q_0 = 2 \pi/r_0$.

 Equations~(\ref{equ_Br_scaling}) and (\ref{equ_Sq_scaling}) can be used to fit numerical or experimental data in order to obtain the roughenss parameters, $\zeta$, $B_0$ and $S_0$. In particular, the roughness exponent $\zeta$ is expected to be a well defined quantity for a given universality class. In the case of fluctuating interfaces this refers to stationary solutions of stochastic differential equations in the thermodynamic limit~\cite{barabasi_surface_growth_95}. However, when using numerical or experimental realizations of interfaces, the measured roughness exponent is expected to fluctuate among those realizations, and thus the importance of properly characterizing fluctuations of roughness parameters.

From a practical point of view, when sufficient statistics can be obtained (with a high resolution, and/or large systems, and/or many systems to average over), fitting the structure factor has been shown to be a generally more reliable method to estimate $\zeta$ than the real-space autocorrelation functions \cite{schmittbuhl_pre_95_reliability_roughness},
essentially because different scaling regimes depending on the length scale are clearly separated in reciprocal space, whereas they are mixed in real space functions.
Moreover, as we discuss below, it can be used to determine roughness exponents ${\zeta>1}$ for super-rough interfaces. More fundamentally, the structure factor is a very important quantity for theoretical developments, and has in particular been shown to be pivotal in the formalism of anomalous scaling~\cite{ramasco_prl_00_generic_scaling}.

Notice that both in numerical and experimental approaches there is usually an intrinsic small length cutoff associated with either discretization of the $z$ direction in the numerical approach or with resolution issues (pixel size) of images in experiments.
This sets on the one hand the lower length scale limit $a$ and the corresponding large wave vector ${q=2 \pi /a}$.
On the other hand the large scale cutoff is given by the system size $L$ and its corresponding small wave vector ${q=2 \pi/L}$.

In addition, since in the present work we will be considering discretization of the $z$ direction, as it is usually the case both in numerical and experimental situations, it is convenient to use a discreteness correction to the wave vector when dealing with the Fourier modes. Consider for instance the discretization as ${z=j \Delta z}$, with ${j=0,1,2,...,L-1}$. Without loss of generality we take ${\Delta z=1}$. Then the interface profile becomes $u_j$ and its Fourier transform is ${\widetilde u_n=L^{-1} \sum_{j=0}^{L-1} u_j e^{-i q_n j}}$, with ${q_n=2 \pi n/L}$.
The large wave vector correction due to the discreteness of $z$ is achieved through the discretization of the Laplacian, ${\nabla^2 u(z) \to u_{j+1} - 2 u_j + u_{j-1}}$, which after Fourier transformation becomes $-q^2 \widetilde u(q) \to \widetilde u_n e^{-i q_n} - 2 \widetilde u_n + \widetilde u_n e^{i q_n}=-4 \sin^2(q_n/2) \widetilde u_n$.
Therefore ${\widetilde q_n = 2 \sin(q_n/2)}$ can be identified as the discretization-corrected wave vector properly controlling small length scale behavior, as it has been explicitly used~\cite{kolton_prl_06_DWdepinning,Rosso_prb_07_numericalFRG, Ferrero2013}.

\subsection{Scaling of ${B(r)}$ for super-rough interfaces}
\label{sec_br_superrough}

Following earlier reports of roughness scaling analyses~\cite{lopez_pre_98_anomalous_scaling,ramasco_prl_00_generic_scaling,Torres_epjb_13_anomalous_scaling}, a generalized formalism distinguishes the global, local, and reciprocal scaling behaviors of the self-affine interface,
characterized by global, local, and reciprocal scaling exponents, respectively.
In this picture, standard Family-Vicsek scaling is given by all three exponents being equal. Any other case falls in one of three categories of anomalous scaling (cf. Ref.~\cite{ramasco_prl_00_generic_scaling} for details). In particular, super-rough interfaces are globally characterized by ${\zeta>1}$, as found using global or reciprocal-space measures, but locally characterized by ${\zeta_{\mathrm{loc}}=1}$, when using a local measure like the roughness function. In this super-rough case Family-Vicsek relations are no longer valid.

Phenomenologically, ${\zeta>1}$ corresponds to the seemingly unphysical case where the transverse fluctuations become unbounded at very large length scales. In such a case, a crossover to a bounded regime may therefore be expected.
Numerically, one-dimensional driven interfaces at the depinning threshold were shown to possess a roughness exponent ${\zeta_{\mathrm{dep}}=1.25}$ when only short-range harmonic contributions to the elastic energy were considered~\cite{Ferrero_pre_13_numerical_exponents}.
Thus, the depinning phase of one-dimensional interfaces should exhibit a crossover from a super-rough regime at small enough length scales to a bounded regime with ${\zeta<1}$.
Experimentally, driven magnetic domain walls were recently shown to exhibit a roughening behavior consistent with this interpretation~\cite{bustingorry_prb_12_depinning, Grassi2018}.
One-dimensional static interfaces are also predicted to exhibit such a crossover at small length scales, at least in a `low-temperature' regime~\cite{agoritsas_2010_PhysRevB_82_184207,agoritsas_2012_FHHtri-numerics}.

To understand the discrepancy between the global and local roughness scaling behavior for super-rough interfaces, the analytical expression of ${B(r)}$ can be considered. The usual derivation starts from the relation between ${B(r)}$ and ${S(q)}$ given in~\eqref{equ_br_sq}.
Assuming a long-time Family-Vicsek scaling for ${S(q)}$ (\eqref{equ_Sq_scaling}), the scaling behavior of ${B(r)}$ is then given in all generality by
\begin{equation}
 B(r) \approx \int_{2\pi/L}^{2 \pi/a} dq \left[1-\cos\left(qr\right)\right] q^{-(1+2\zeta)}.
\label{equ_br_general}
\end{equation}
When ${0 < \zeta \leq 1}$ and taking the limits ${a \to 0}$ and ${L \to \infty}$ the integral converges and the Family-Vicsek scaling relation is recovered, with a single $\zeta$ value describing both the local and global correlation functions. In the case ${\zeta > 1}$, taking ${a \to 0}$ and for large but finite values of $L$, the roughness function ${B(r)}$ obeys the general scaling behavior for `super-rough' interfaces

\begin{equation}
 B(r)
 \approx r^2\left[ -A_0r^{2(\zeta-1)} + C(L) \right],
\label{equ_br_superrough}
\end{equation}

\noindent where $\zeta$ is the reciprocal-space roughness exponent,  $C(L) = A_1 L^{2(\zeta-1)}$ is an $L$-dependent constant, and  $A_0$ and $A_1$ are positive constants.
The presence of the $r^2$ prefactor indicates that, when taking the limit ${L\to\infty}$ first and then the large $r$ limit, the local roughness exponent saturates to ${\zeta_{\mathrm{loc}} = 1}$.
We note that this expression is slightly more general than the one reported in Ref.~\cite{lopez_pre_98_anomalous_scaling}, which holds only for large values of $L$. Equation~(\ref{equ_br_superrough}) can be rewritten in the form
\begin{equation}
 C(L) r^2 - B(r) = A_0 r^{2 \zeta}.
\end{equation}
This power-law behavior can be used to obtain the roughness exponent $\zeta>1$ and the amplitude $A_0$ using a local measure. Such a super-rough behavior will be illustrated in Sec.~\ref{sec_num_dep} using numerical simulations.

\subsection{The relevance of statistical averaging}

As the scaling relations in Eqs.~(\ref{equ_Br_scaling}) and (\ref{equ_Sq_scaling}) only hold with the appropriate statistical averaging, a crucial step in roughness analysis of experimental interfaces is to assess the minimal number of independent configurations necessary to achieve a meaningful estimation of the roughness parameters~\cite{Jordan2020}. Furthermore, one may ask how representative of the actual roughness exponent, the one characterizing a universality class, is the value obtained from a single measurement. In both cases, we can expect the answer to be both size and method-dependent.

For numerical simulations where a large number of independent realizations can be available, a meaningful estimation of the roughness parameters can readily be obtained by computing the desired correlation functions averaged over the number of realizations, and subsequently fitting single $\zeta$ and $B_0$ values from the power-law behavior. In contrast, for experimentally imaged interfaces the amount of different realizations is typically small and may in addition suffer from differences in size and/or resolution. Therefore, a common practice is to compute the roughness function of a single interface, $B_i(r)$, and extract individual roughness parameters $\zeta_i$ and ${B_0}_i$ of a single interface profile using $B_i(r) = {B_0}_i (r/r_0)^{2 \zeta_i}$. Mean values are then expected to be representative of the scaling properties:
\begin{eqnarray}
\overline{\zeta} = \frac{1}{N} \sum_{i=1}^{N} \zeta_i,\\
\overline{B_0} = \frac{1}{N} \sum_{i=1}^{N} {B_0}_i,
\end{eqnarray}
where $N$ is the number of independent measurements.
The same procedure can be followed to obtain $\zeta_i$ and ${S_0}_i$ from individual structure factors $S_i(q)$ to get mean values $\overline{\zeta}$ and $\overline{S_0}$.
It is important to note that $\overline{\zeta}$ is not necessarily equivalent to $\zeta$ unless the underlying distribution for the roughness exponent happens to be symmetric. Therefore, the skewness of the $\zeta_i$ histogram is indicative of the validity of this method and the accuracy of $\overline{\zeta}$~\cite{Jordan2020, guyonnet_prl_12_multiscaling}.

\section{Numerical simulations}
\label{sec_numerical}

We shall use three different numerical models to evaluate statistical fluctuations of roughness parameters: \textit{(i)}~an elastic line model subject to thermal fluctuations, belonging to the Edwards-Wilkinson universality class, \textit{(ii)}~a directed polymer in equilibrium within a quenched disordered environment (equilibrium quenched-Edwards-Wilkinson universality class), and \textit{(iii)}~a driven elastic line in a quenched disorder exactly at the depinning critical point, within the depinning quenched-Edwards-Wilkinson universality class.
Note that these three models have already been benchmarked on their own in previous studies, as we cite accordingly when we introduce them thereafter.

\subsection{Elastic line with thermal noise}
\label{sec_num_th}

Fluctuations of an interface subjected to a thermal noise are well described through the time evolution of the interface profile ${u(z,t)}$ given by the Edwards-Wilkinson equation
\begin{equation}
 \label{equ_EW}
 \partial_t u(z,t) = c \left[ u(z+1,t) - 2 u(z,t) + u(z-1,t) \right] + \eta(z,t),
\end{equation}
where ${u(z,t)}$ is the time-dependent position of the interface with elasticity $c$ and $\eta(z,t)$ is a white noise representing contact with a thermal bath of intensity $T$ with $\langle \eta(z,t) \rangle = 0$ and $\langle \eta(z,t) \eta(z',t') \rangle = 2 T \delta_{zz'} \delta(t-t')$, where $\delta_{zz'}$ and $\delta(t)$ are respectively Kronecker and Dirac \emph{delta} functions. For simplicity, in this equation we have used a discrete variable along the interface, \textit{i.e.}~$z$ takes discrete values, and a continuum variable for the displacement field $u$. Taking as initial condition $u(z,t=0) = 0$, \eqref{equ_EW} is solved using periodic boundary conditions with discrete time units, $\delta t = 0.05$, and using $T=1$ and $c = 1/2$. Dynamics under a white noise makes the interface to roughen, with its roughness increasing with time until a correlation length characteristic of the fluctuations reaches the size of the system, $\xi(t) \approx L$. After that, the system dynamics reaches a stationary limit and the roughness fluctuates around a size-dependent mean value, which is analytically known to be ${B(r)=\frac{Tr}{c} (1-r/L)}$ for a continuous interface. This state is representative of thermal roughness of a fluctuating interface. We take $N = 1000$ independent realizations of profiles $u(z)$ in such state for different system size values, $L = 256$, $512$ and $1024$.

\begin{figure}
\begin{center}
\includegraphics[width=0.45\columnwidth]{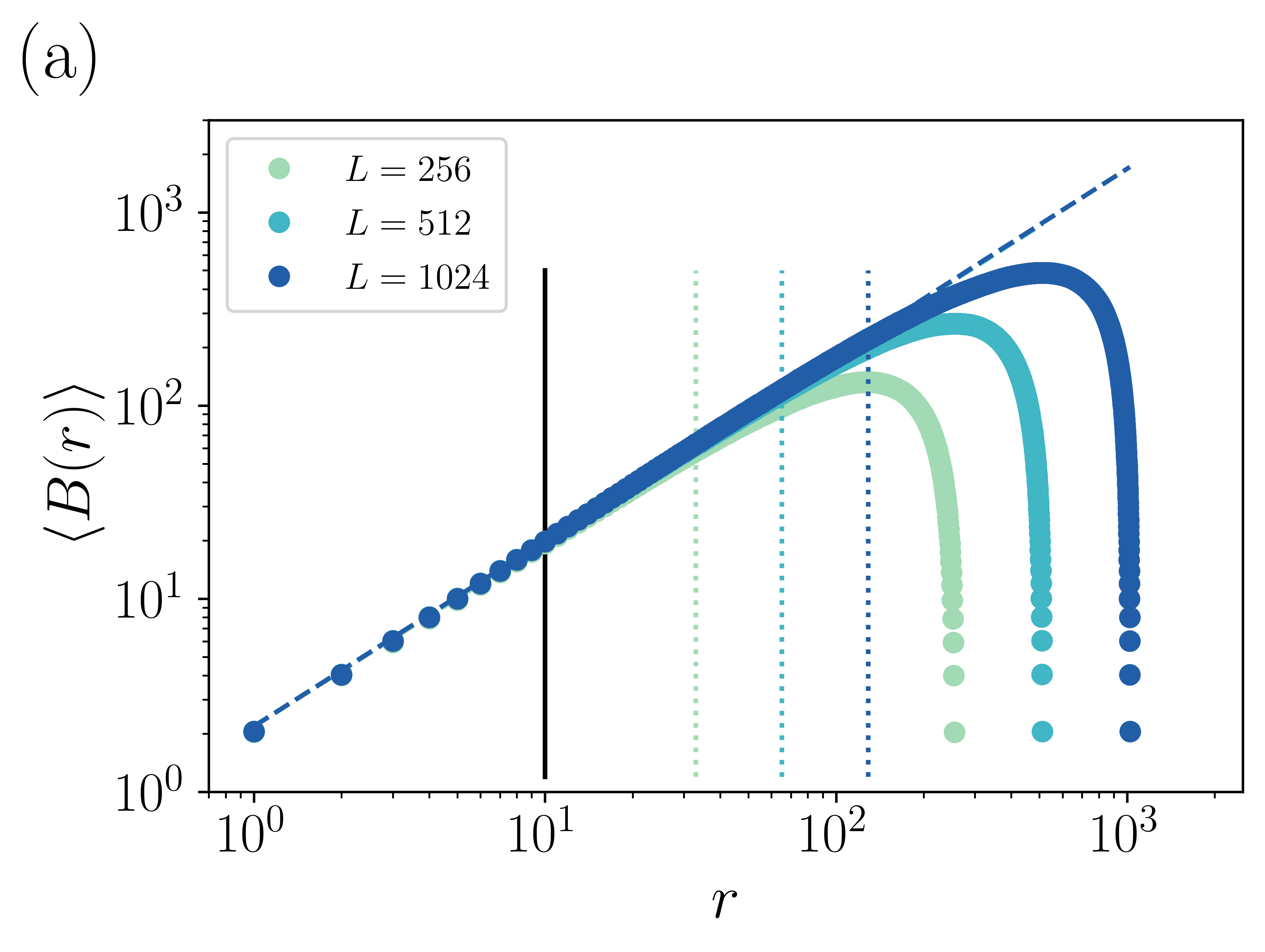}
\includegraphics[width=0.45\columnwidth]{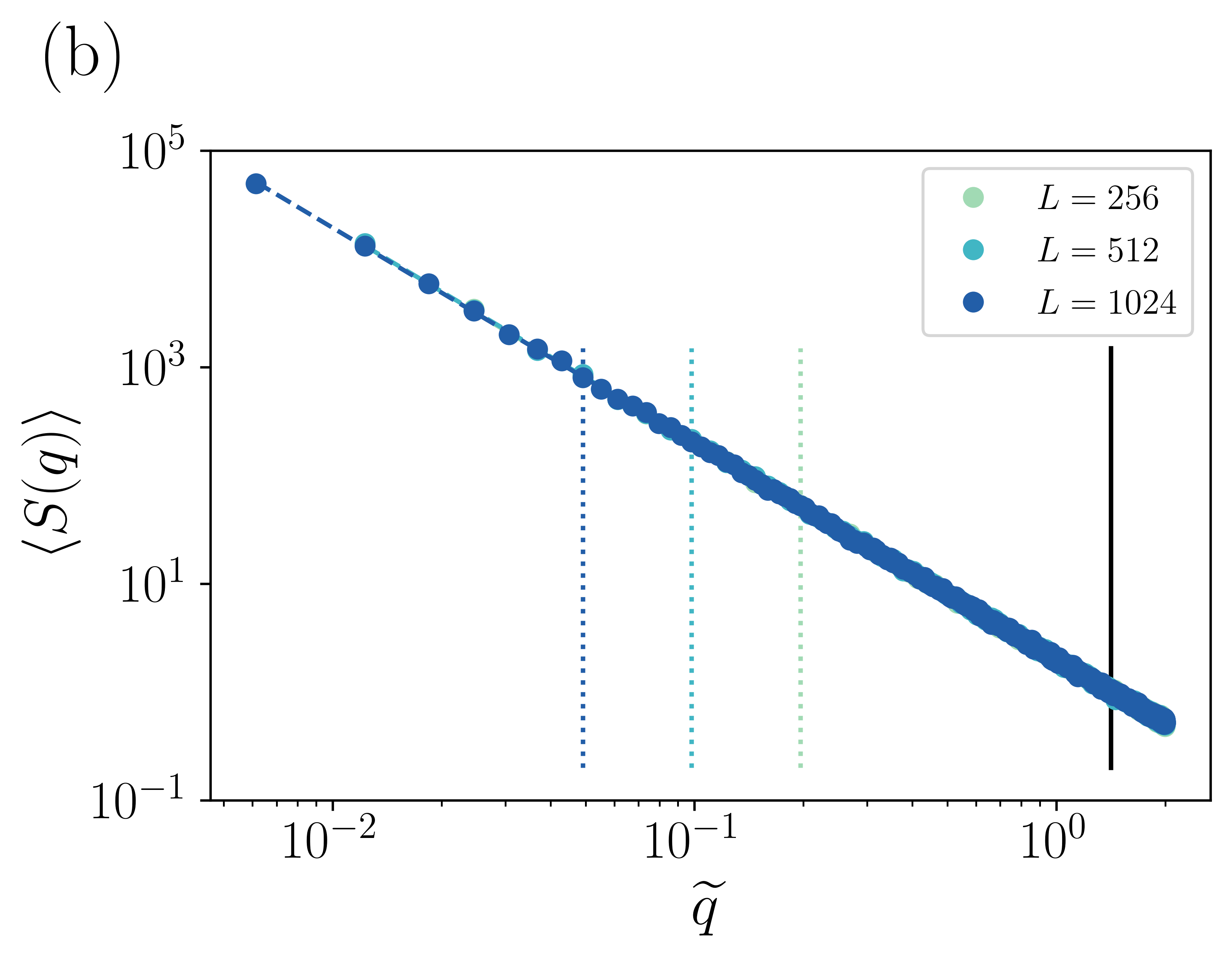}
\end{center}
\caption{
(a)~ Averaged displacement-displacement autocorrelation function ${\langle B(r) \rangle}$ and (b) averaged structure factor ${\langle S(q) \rangle}$ for the elastic line with thermal noise model.
The solid vertical lines indicate the common small length scale bound of the fitting ranges, while the large length scale bounds depend on the system size and are indicated by the vertical dotted lines.
Analytically we expect to have ${B(r)=\frac{Tr}{c}(1-r/L)}$, for the periodic boundary conditions we consider, hence the observed parabola in a log-log scale in (a).
}
\label{fig_corr_ew}
\end{figure}

Figure~\ref{fig_corr_ew}(a) shows the averaged roughness function ${\langle B(r) \rangle}$ for different system sizes, as indicated. Power-law behavior is observed for small $r$ values. Then a maximum of ${\langle B(r) \rangle}$ is reached at size-dependent values with a drop of the roughness function due to periodicity of the interfaces. From the power-law region at intermediate length scales we can fit roughness parameters, $\zeta$ and $B_0$, according to \eqref{equ_Br_scaling}. The power-law behavior breaks down around ${r=L/2}$, corresponding to the local maximum of  ${\langle B(r) \rangle}$ observed in \figref{fig_corr_ew}(a). The most adequate fitting ranges are found to extend between ${r=10}$ and ${r \approx L/8}$, indicated by the vertical solid and dotted lines in \figref{fig_corr_ew}(a).
Fitted values are shown in the figure and reported in Table~\ref{tab_ew}, with error bars obtained from the power-law fitting. These error bars are rather small and do not take into account variations of the fitting range~\cite{Jordan2020}. The roughness exponent is close to the expected $\zeta_{\mathrm{th}} = 1/2$ value, but always slightly underestimated. Because we know in this case the exact analytical prediction for ${B(r)}$ and it is an inverted parabola, this underestimation can be directly attributed to the periodicity of the interface and will always be there for any fitting range we choose.

The  average structure factor  ${\langle S(q) \rangle}$ is presented in \figref{fig_corr_ew}(b). The data presents power-law behavior in all the studied range and the roughness parameters, $\zeta$ and $S_0$, can be extracted fitting the data using \eqref{equ_Sq_scaling}. In this case, the lower bounds for the fitting range are size-dependent and we used $2\pi/(L/8)$, vertical dotted lines in \figref{fig_corr_ew}(b). The fitting range can be extended in comparison to the  ${\langle B(r) \rangle}$ and we used the same $\pi/2$ value for all system sizes (vertical solid line). We have also used the wave vector number corrected by discretization effects, ${\widetilde q_n = 2 \sin(q_n/2)}$.
Values for the roughness parameters are reported in Table~\ref{tab_ew} and good agreement with the expected values are found.

The roughness parameters obtained from the averaged correlation functions are in good agreement with the expected values. However, experimental observations are generally based on a few independent realizations of interfaces, not even using a fixed system size. Therefore, we compute here roughness parameters for individual independent realizations of fluctuating interfaces to obtain information about the spread of the data, useful to interpret experimental results.

\begin{figure}
\begin{center}
\includegraphics[width=0.48\columnwidth]{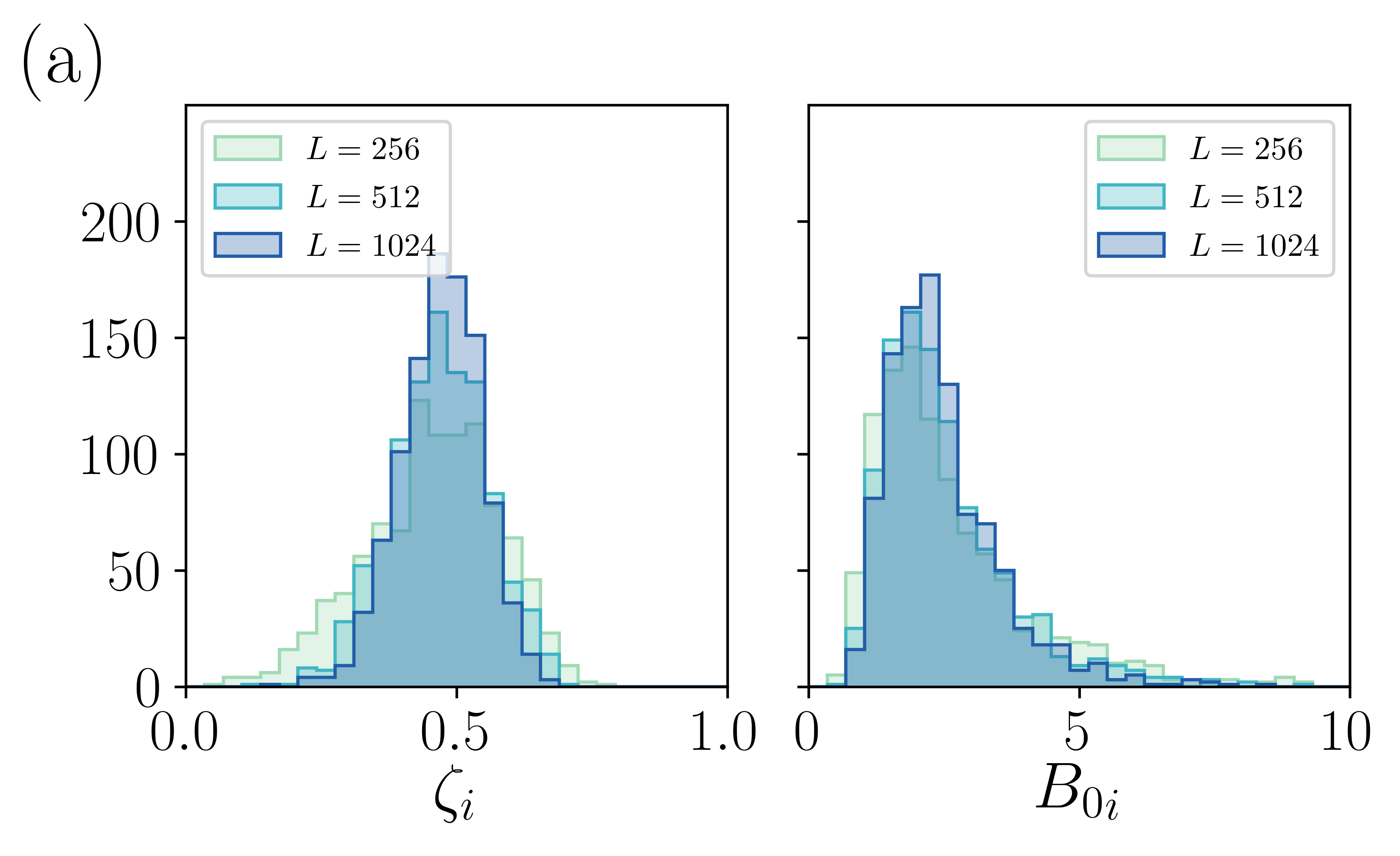}
\includegraphics[width=0.48\columnwidth]{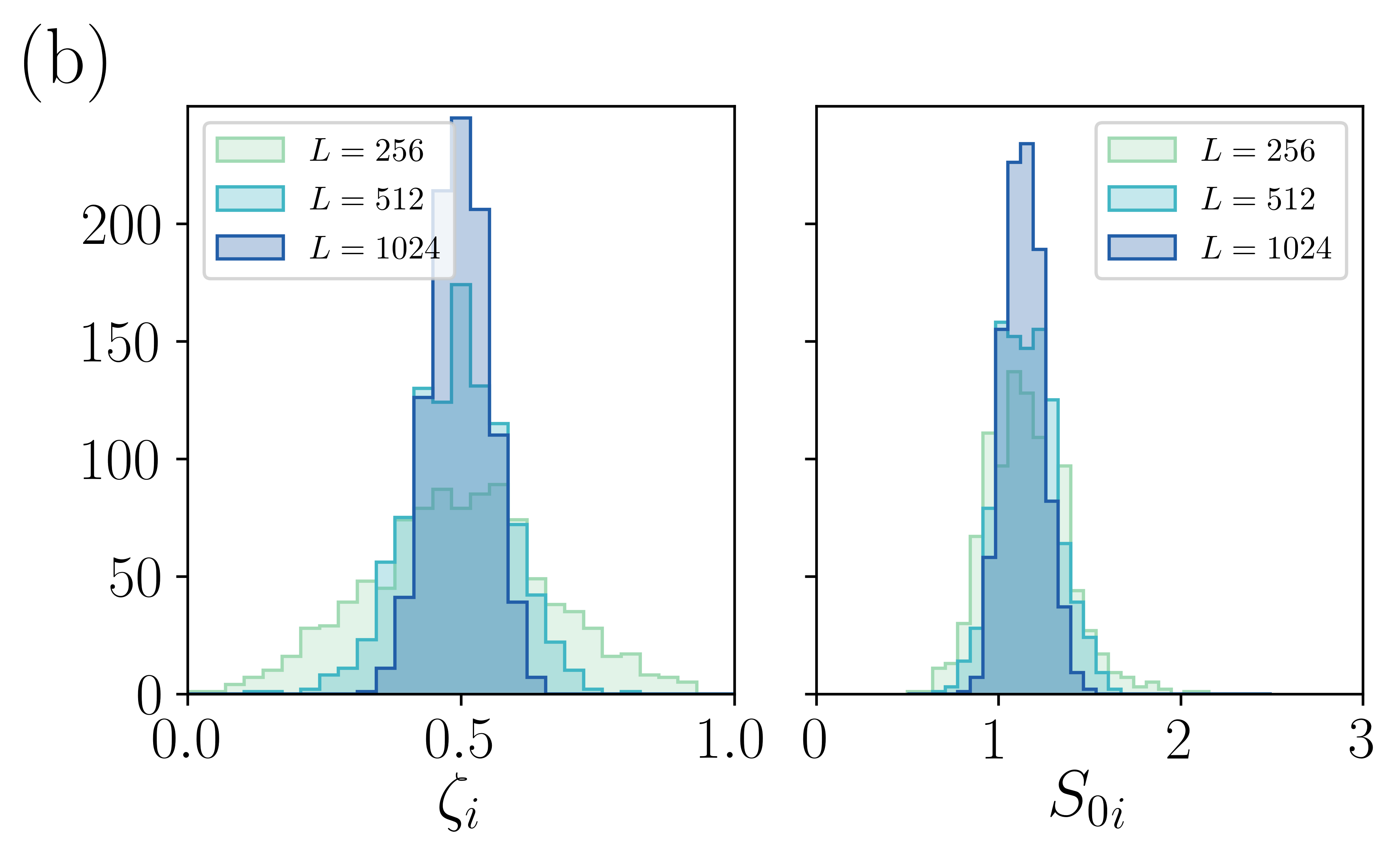}
\end{center}
\caption{
Histograms of the roughness exponent obtained using (a)~${B(r)}$ and (b)~${S(q)}$ for individual numerical interfaces corresponding to elastic lines with thermal noise.
The mean value and the variance of each distribution are indicated in Table~\ref{tab_ew}.
}
\label{fig_dist_ew}
\end{figure}

The roughness function and structure factor are obtained from each interface profile and then are fit to power-law behavior. The distribution of the resulting roughness parameters obtained using the roughness function and structure factor are presented as histograms in Fig.~\ref{fig_dist_ew}(a) and Fig.~\ref{fig_dist_ew}(b), respectively. Results for different system sizes are shown, as indicated.
Mean values $\overline{\zeta}$, $\overline{B_0}$, and $\overline{S_0}$, with their error bars computed as the variance of the distributions divided by the square root of the number of realizations, $\sigma/\sqrt{N}$, are presented in Table~\ref{tab_ew}.
We observe in Fig.~\ref{fig_dist_ew} that the distributions of the roughness exponents are considerably broad in general. Although the mean value of the distributions, $\overline{\zeta}$, is very close to the expected value $\zeta_{\mathrm{th}} =1/2$, values are spread in a finite range inside the $0 < \zeta_i < 1$ window. In general, the widths of the distributions decrease with system size, as expected. In addition, ${B_0}_i$ and ${S_0}_i$ values are also broadly distributed. In particular, the distribution of ${B_0}_i$ are considerable more broad and present an appreciably skewness, while the distributions of $\zeta_i$ and ${S_0}_i$ are symmetric.

\begin{table}[!th]
\begin{center}
\begin{tabular}{|c|c|c|c|c|}
\multicolumn{5}{c}{Elastic line with thermal noise ($\zeta_{\mathrm{th}} = 1/2$)}	\\
\hline
\multicolumn{2}{|c|}{$L$} & 256 & 512 & 1024 \\
\hline\hline
$\langle B(r) \rangle$ & $\zeta$ & $0.4723 \pm 0.0008$ & $0.480 \pm 0.001$ & $0.4810 \pm 0.0009$ \\
& $B_0$ & $2.192 \pm 0.006$ & $2.18 \pm 0.01$ & $2.182 \pm 0.009$ \\
\hline
$\langle S(q) \rangle$ & $\zeta$ & $0.497 \pm 0.005$ & $0.498 \pm 0.003$ & $0.496 \pm 0.002$ \\
& $S_0$ & $2.008 \pm 0.009$ & $2.038 \pm 0.007$ & $2.022 \pm 0.005$ \\
\hline
$B_i(r)$ & $\overline{\zeta}$ & $0.456 \pm 0.004$ & $0.468 \pm 0.003$ & $0.471 \pm 0.002$ \\
& $\overline{B_0}$ & $2.69 \pm 0.06$ & $2.52 \pm 0.04$ & $2.47 \pm 0.03$ \\
\hline
$S_i(q)$ & $\overline{\zeta}$ & $0.494 \pm 0.005$ & $0.494 \pm 0.003$ & $0.497 \pm 0.002$ \\
& $\overline{S_0}$ & $1.153 \pm 0.007$ & $1.157 \pm 0.005$ & $1.138 \pm 0.003$ \\
\hline
\end{tabular}
\end{center}
\caption{
Roughness parameters for the elastic line with thermal noise. Reported values for $\zeta$, $B_0$ and $S_0$ are obtained by fitting power-law behavior to average roughness function and structure factor, $\langle B(r) \rangle$ and $\langle S(q) \rangle$, respectively. $\overline{\zeta}$, $\overline{B_0}$ and $\overline{S_0}$ are the mean values of the distributions of individual values for each interface.
\label{tab_ew}
}
\end{table}

\subsection{Directed polymer in equilibrium with quenched disorder}
\label{sec_num_equ}

Interfaces in equilibrium, within an heterogeneous substrate, can be generated by allowing a directed polymer living in a disordered energy landscape to relax to its minimum energy configuration. One-dimensional equilibrated interfaces in weak collective random-bond disorder were simulated from a directed polymer model~\cite{mezard_jdpi_92_manifolds} on a discretized square lattice.
The position of the polymer is given by ${u(z)}$, with $u$ and ${z \geq 0}$ taking discrete values, mimicking a fluctuating interface. The solid-on-solid restriction ${|u(z+1) - u(z)| = \pm 1}$ provides the effective short-range elasticity to the polymer model. An uncorrelated Gaussian random potential distributed on each lattice site, ${V(u,z)}$, is used to model a disordered energy landscape. Disordered potential correlations are given by ${\overline{V(u,z)V(u',z')} = D \, \delta_{u,u'}\delta_{z,z'}}$, where $D$ is the strength of the disorder. The equilibrium zero temperature configuration was obtained using the transfer-matrix method~\cite{Kardar_prl_85_DP} with a droplet geometry, \textit{i.e.}~with one end pinned at the origin while the other end is free.
Given the disordered potential ${V(u,z)}$, the probability weight ${Z(u,z)}$ of a polymer starting at ${(0,0)}$ and ending at ${(u,z)}$ is given recursively by
\begin{equation}
 Z(u,z) = e^{-\beta V(u,z)} \left[ Z(u-1,z-1) + Z(u+1,z-1) \right],
\end{equation}
with initial condition ${Z(u,0) = \delta_{u,0}}$, and $\beta$ the inverse temperature parameter. For each realization of the disordered potential the path of minimum energy corresponds to the largest weight, thus defining equilibrium interface~\cite{Kardar_prl_85_DP}.
In this canonical case, the value of the roughness exponent ${\zeta_{\mathrm{eq}} = 2/3}$ is already well known~\cite{huse_henley_fisher_1985_PhysRevLett55_2924}.
For the numerical simulations systems with $L=512$, $1024$ and $2048$ sites were used, with ${N = 1000}$ different disorder realizations for each size.

\begin{figure}
\begin{center}
\includegraphics[width=0.45\columnwidth]{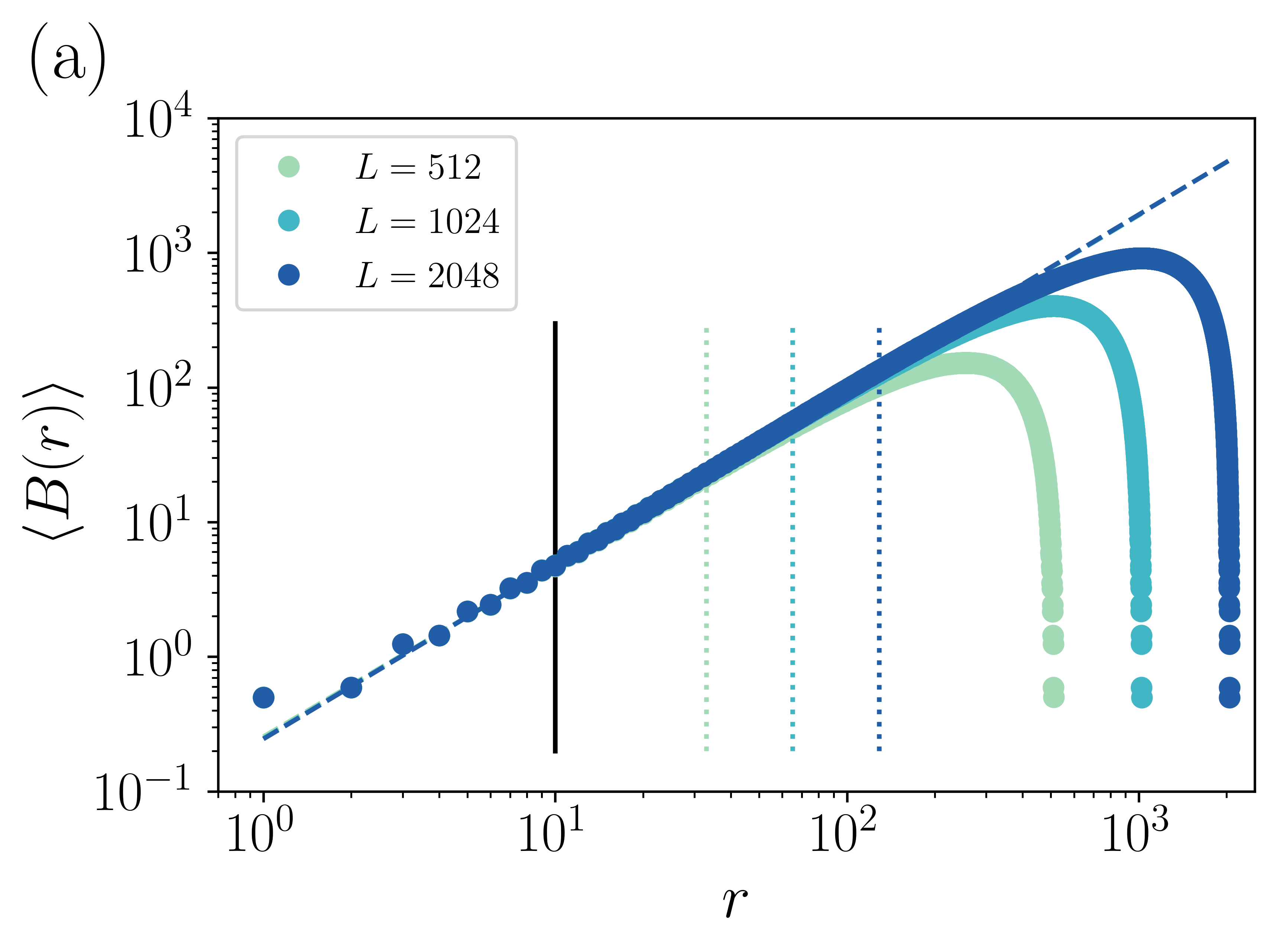}
\includegraphics[width=0.45\columnwidth]{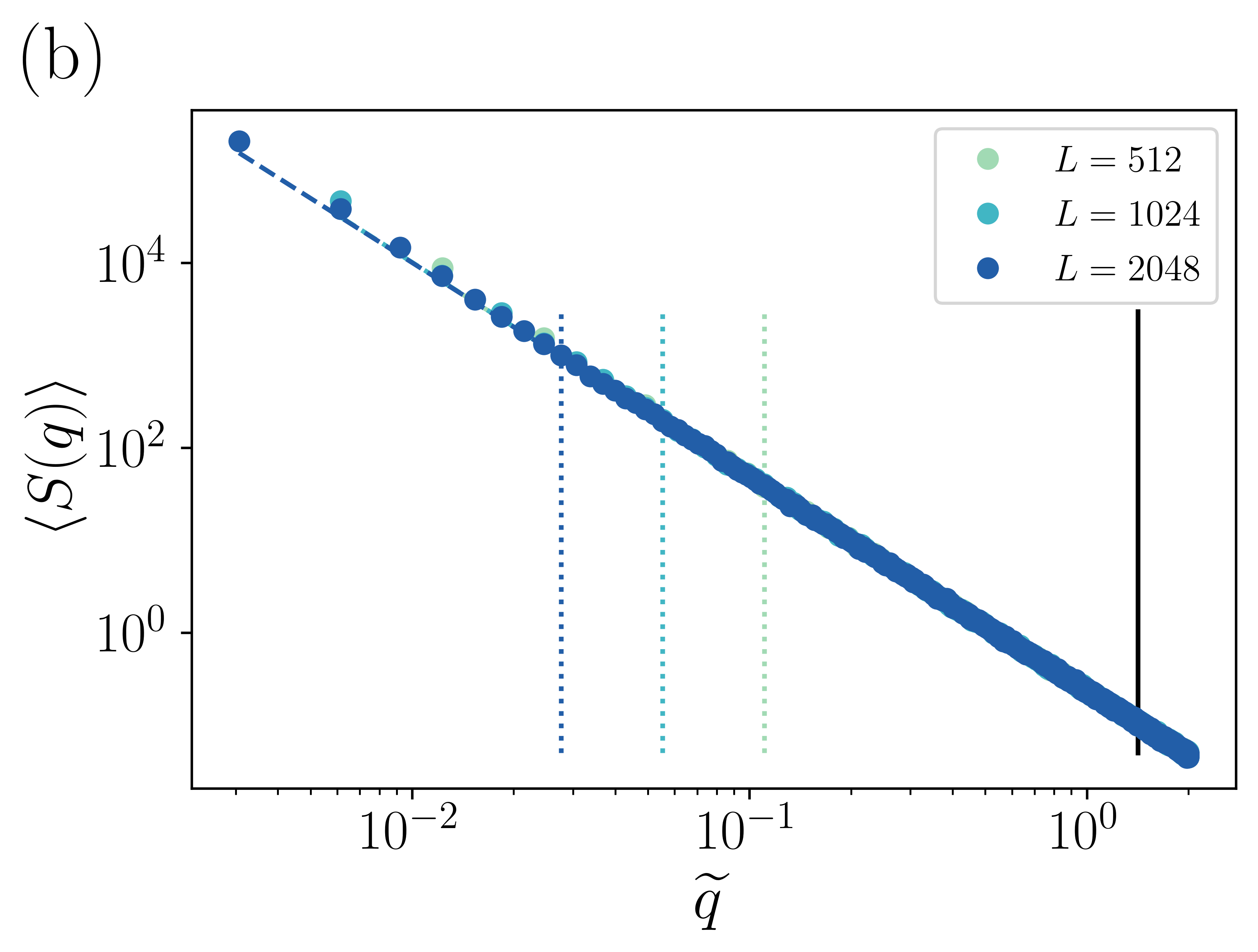}
\end{center}
\caption{
(a) Average displacement-displacement autocorrelation function ${\langle B(r) \rangle}$ and (b) average structure factor ${\langle S(q) \rangle}$ for the directed polymer in equilibrium with a disordered environment.
The solid vertical lines indicate the common small length scale bound of the fitting ranges, while the large length scale bounds depend on the system size and are indicated by the vertical dotted lines.
}
\label{fig_corr_eq}
\end{figure}

Figures~\ref{fig_corr_eq}(a) and ~\ref{fig_corr_eq}(b) show the averaged roughness function ${\langle B(r) \rangle}$ and structure factor ${\langle S(q) \rangle}$, respectively, for different system sizes. The roughness functions were fitted in the range $10 < r < L/8$, while structure factor were fitted for values in $2\pi/(L/8) < q < \pi/2$, with vertical lines indicating the fitting range, as described in the caption. The obtained roughness parameters are summarized in Table~\ref{tab_eq}.
The roughness exponents are close to the expected value ${\zeta_{\mathrm{eq}} = 2/3}$. Error bars from the power-law fitting underestimate sample-to-sample fluctuations.
For the roughness function, systematic errors due to finite system size are due to the finite size of the fitting range and therefore appear to lead to slightly underestimated values of the roughness exponent (as it was already the case in Table~\ref{tab_ew}). We emphasize here that for real systems presenting both experimental artifacts and fewer realizations for averaging, we expect this effect to be significantly greater.
The structure factor functions presented in \figref{fig_corr_eq}(b) show again less finite-size effects when compared with the roughness function, albeit with the same slight underestimating trend.

\begin{figure}
\begin{center}
\includegraphics[width=0.48\columnwidth]{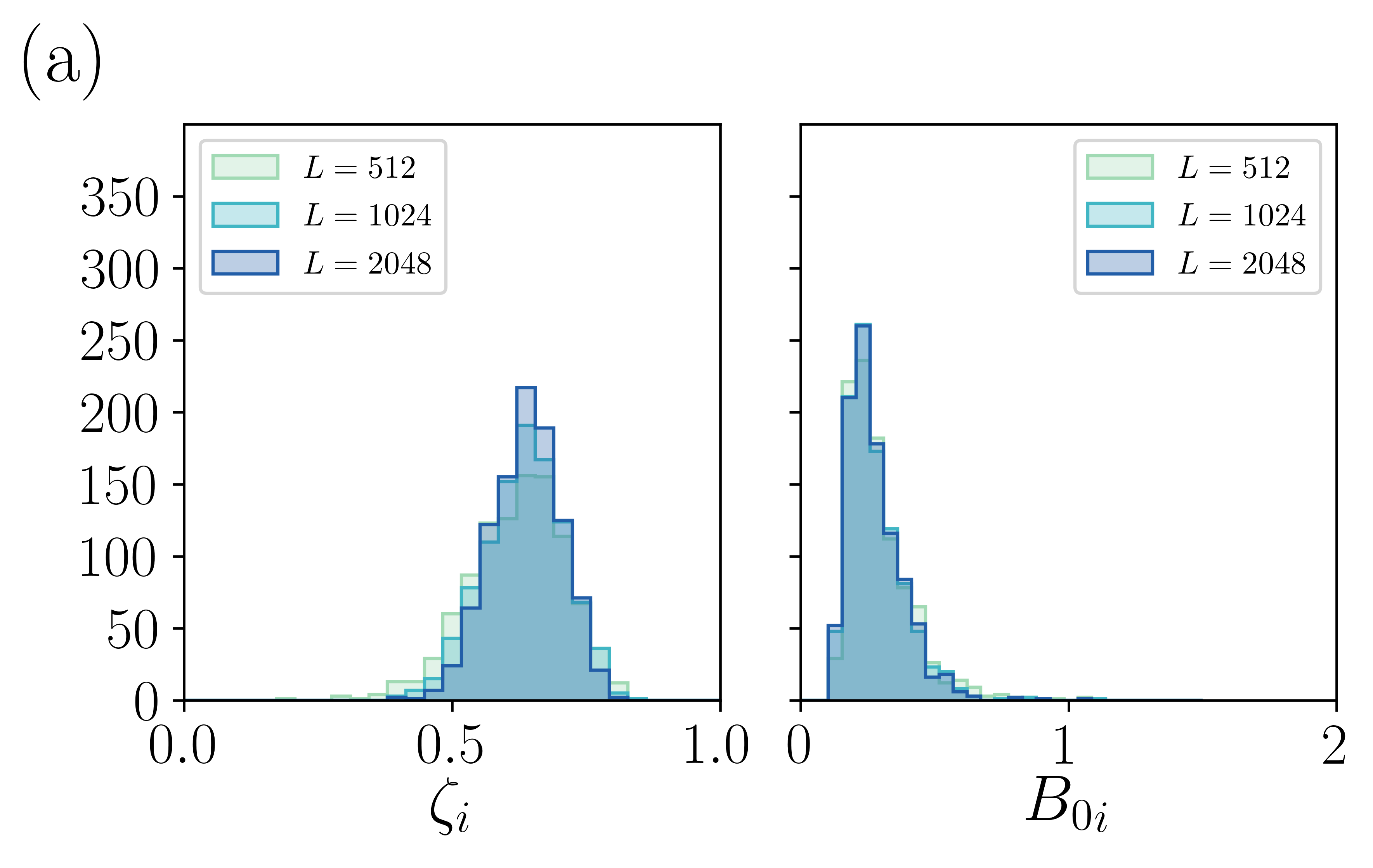}
\includegraphics[width=0.48\columnwidth]{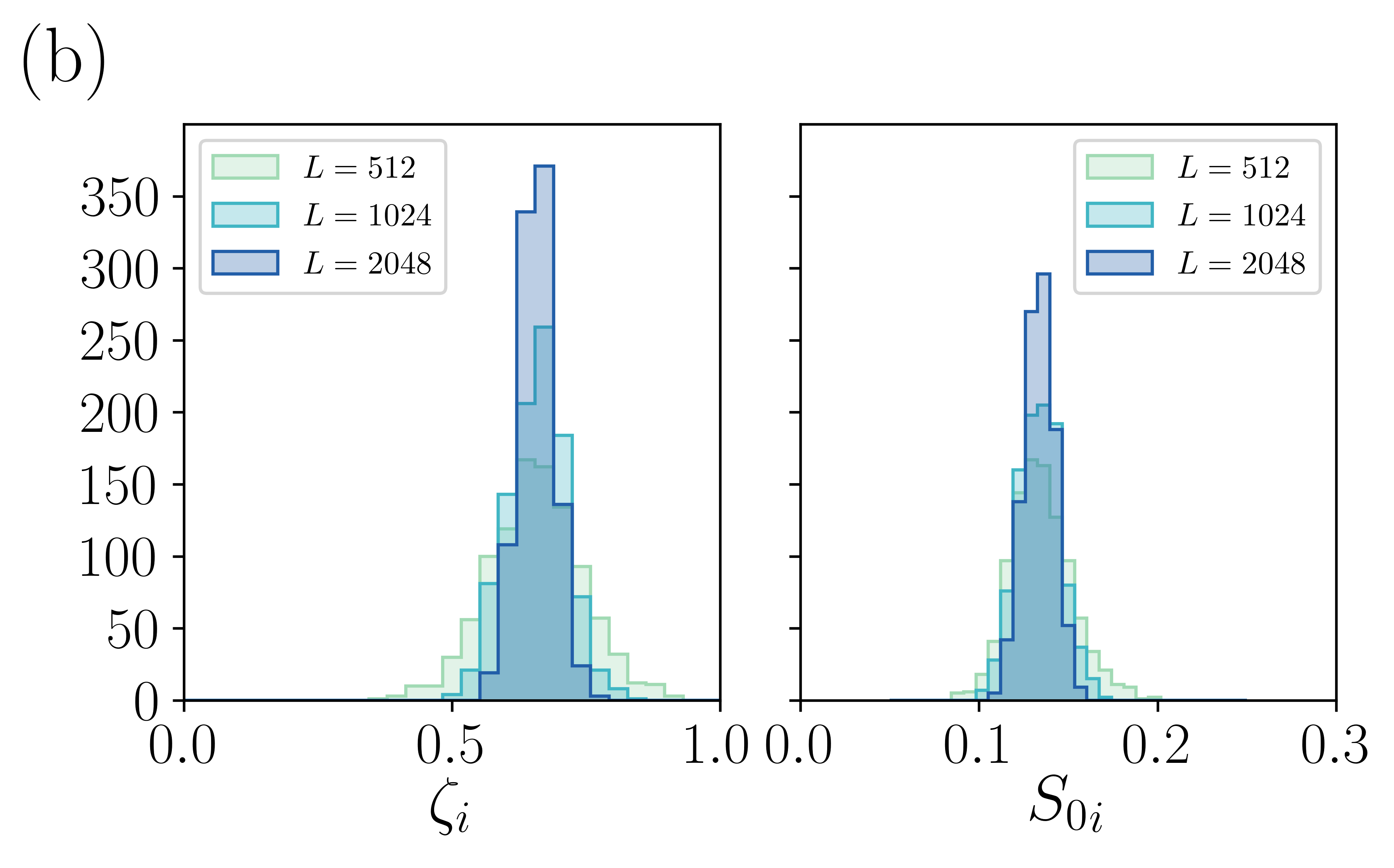}
\end{center}
\caption{
Histograms of the roughness exponent obtained using (a)~${B(r)}$ and (b)~${S(q)}$ for individual numerical interfaces corresponding to elastic lines in equilibrium with a disordered environment.
The mean value and the variance of each distribution are indicated in Table~\ref{tab_eq}.
}
\label{fig_dist_eq}
\end{figure}

\begin{table}[!th]
\begin{center}
\begin{tabular}{|c|c|c|c|c|}
\multicolumn{5}{c}{Equilibrium elastic line with quenched disorder ($\zeta_{\mathrm{eq}} = 2/3$)}	\\
\hline
\multicolumn{2}{|c|}{$L$} & 512 & 1024 & 2048 \\
\hline\hline
$\langle B(r) \rangle$ & $\zeta$ & $0.638 \pm 0.006$ & $0.647 \pm 0.002$ & $0.6490 \pm 0.0007$ \\
& $B_0$ & $0.255 \pm 0.006$ & $0.248 \pm 0.002$ & $0.245 \pm 0.001$ \\
\hline
$\langle S(q) \rangle$ & $\zeta$ & $0.657 \pm 0.003$ & $0.658 \pm 0.002$ & $0.657 \pm 0.001$ \\
& $S_0$ & $0.2389 \pm 0.0007$ & $0.2369 \pm 0.0006$ & $0.2376 \pm 0.0004$ \\
\hline
$B_i(r)$ & $\overline{\zeta}$ & $0.621 \pm 0.003$ & $0.634 \pm 0.002$ & $0.638 \pm 0.002$ \\
& $\overline{B_0}$ & $0.295 \pm 0.004$ & $0.281 \pm 0.003$ & $0.277 \pm 0.003$ \\
\hline
$S_i(q)$ & $\overline{\zeta}$ & $0.656 \pm 0.003$ & $0.658 \pm 0.002$ & $0.657 \pm 0.001$ \\
& $\overline{S_0}$ & $0.1349 \pm 0.0005$ & $0.1338 \pm 0.0004$ & $0.1336 \pm 0.003$ \\
\hline
\end{tabular}
\end{center}
\caption{
Roughness parameters for the equilibrium elastic line with quenched disorder. Reported values for $\zeta$, $B_0$ and $S_0$ are obtained by fitting power-law behavior to average roughness function and structure factor, $\langle B(r) \rangle$ and $\langle S(q) \rangle$, respectively. $\overline{\zeta}$, $\overline{B_0}$ and $\overline{S_0}$ are the mean values of the distributions of individual values for each interface.
We considered ${N=1000}$ independent disorder realizations.
\label{tab_eq}
}
\end{table}

The different sensitivities to size effects observed in real-space and reciprocal-space methods, only marginally observable on quantities averaged over $1000$ different disorder realizations, can be expected to become much more prominent in studies where disorder averaging is significantly reduced. This is immediately verified, as can be seen from the distributions of individual realization exponents extracted from ${B(r)}$ and ${S(q)}$, shown in \figref{fig_dist_eq}.
In both cases, scaling exponents are obtained from power-law fits for each single interface, with the same fitting regions as the ones defined for the averaged quantities. For ${L=2048}$, the histogram of the individual exponent values constructed from ${B(r)}$ appears wider than the one from ${S(q)}$, with standard deviations of 0.06 and 0.03, respectively.
In contrast, both methods yield histograms of comparable widths for ${L=512}$, suggesting the convergence of the distribution with increasing system size happens faster for the structure factor.
Another notable feature of the distributions presented in \figref{fig_dist_eq} is the slight negative skewness of all histograms, also decreasing with increasing system size, but significantly more pronounced for ${B(r)}$. For the roughness exponent, this can be attributed to the inherent ${\zeta<1}$ cutoff of the method, effectively compressing the histogram to the right.
For ${B_0}_i$ histograms, the small but non-negligible skewness is correlated with a slight underestimation of the roughness exponent.
Distributions are broader and with a larger skewness in the case of ${B_0}_i$ than ${S_0}_i$

\subsection{Driven elastic line at critical depinning}
\label{sec_num_dep}

When an interface living in a disordered energy landscape is driven by an external force, its zero temperature critical depinning state corresponds to the configuration encountered exactly at the depinning force, separating zero velocity from finite velocity steady states~\cite{Ferrero2013}. This critical depinning state then results from the interplay between the elasticity of the interface, the disordered energy landscape and the external force.
We use a simple model to describe the dynamics of an elastic interface in a disordered energy landscape given by the quenched Edwards-Wilkinson equation at zero temperature,
\begin{eqnarray}
 \partial_t u(z,t) = & &u(z+1,t) - 2 u(z,t) + u(z-1,t) \nonumber \\
  &+& F_p\left( u(z,t),z \right) + F,
\end{eqnarray}
where ${u(z,t)}$ is the time-dependent position of the interface, $F$ is an homogeneous external drive, and ${F_p(u,z)= - \partial_u V(u,z)}$ is a pinning force. The disordered potential ${V(u,z)}$ has zero mean and correlations ${\overline{\left[ V(u,z)-V(u',z') \right]^2} = D \delta_{z,z'} R(u-u')}$, with ${R(u)}$ decaying in a finite range. For this model, there exists a finite force value $F_d$ separating pinning configurations for ${F < F_d}$ from moving configurations for ${F > F_d}$. $F_d$ is the depinning force and Middleton theorems~\cite{Middleton1992} assure that a unique critical depinning configuration ${u_c(z)}$ exists, corresponding to ${\partial_t u(z,t) = 0}$ at $F_d$. 
One-dimensional interfaces in a critical depinning state were obtained using the algorithm developed in Ref.~\cite{rosso_prl_03_depinning}, where the interface is forced to its last zero-velocity state under a finite driving force. The roughness exponent characterizing critical depinning interfaces is then ${\zeta_{\mathrm{dep}}=1.25}$ (see Ref.~\cite{Ferrero_pre_13_numerical_exponents}).
The simulation box longitudinal and transverse sizes $L$ and $M$ were chosen such that ${L/M \sim 3-10}$ in order to avoid spurious effects due to periodic boundary conditions~\cite{Bustingorry_prb_10_random_periodic,Bustingorry_pip_10_random_periodic,Kolton_jstat_13_fc_random_periodic}, and simulations were performed with $L=256$, $512$, and $1024$, with $N = 1000$ independent disorder configurations for each size. In this model, the internal coordinate of the interface position is a discrete variable ${z=1,2,\ldots,L}$ and the transverse coordinate is a continuous variable, as for the elastic line without disorder described in Sec.~\ref{sec_num_th}.

\begin{figure}
\begin{center}
\includegraphics[width=0.45\columnwidth]{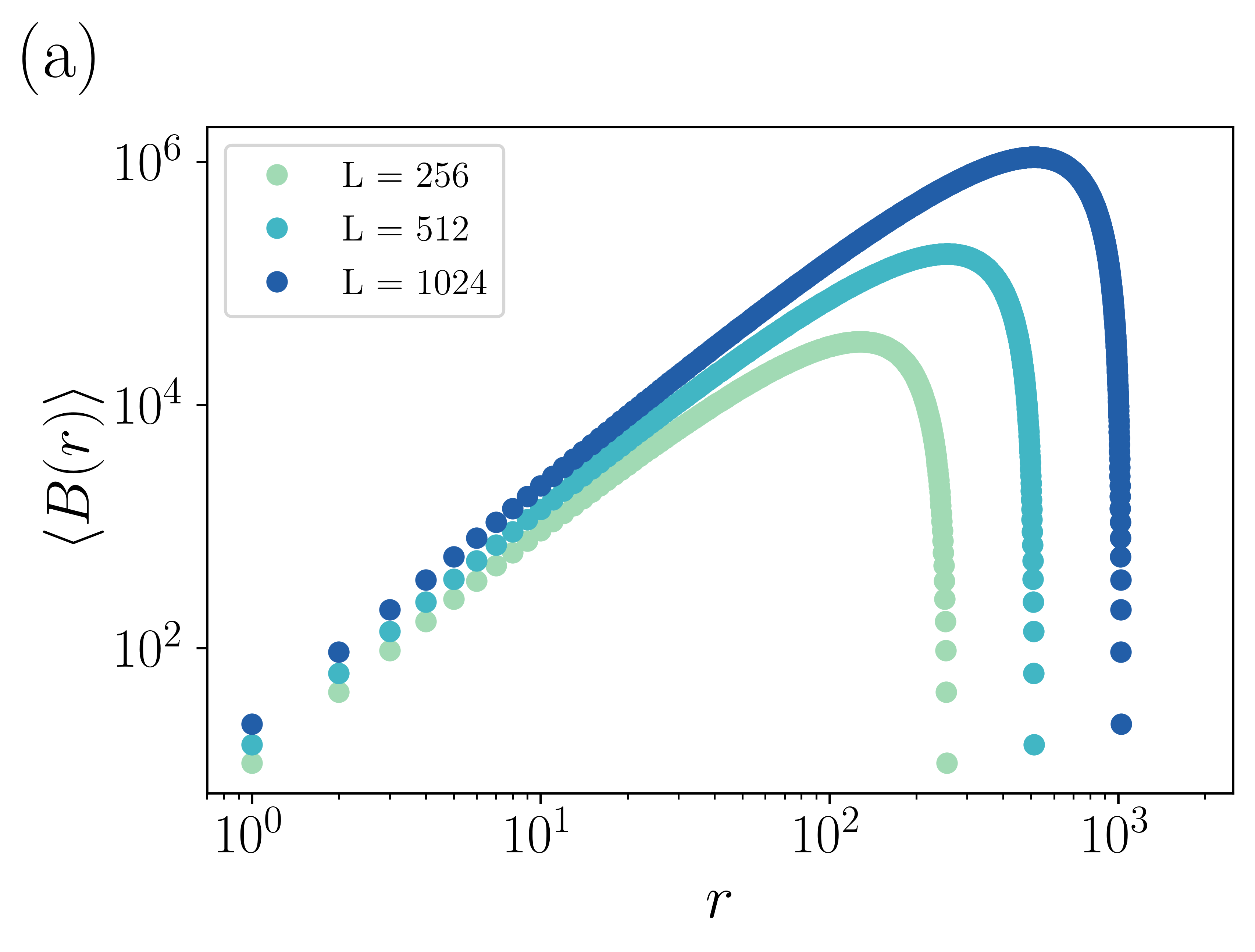}
\includegraphics[width=0.45\columnwidth]{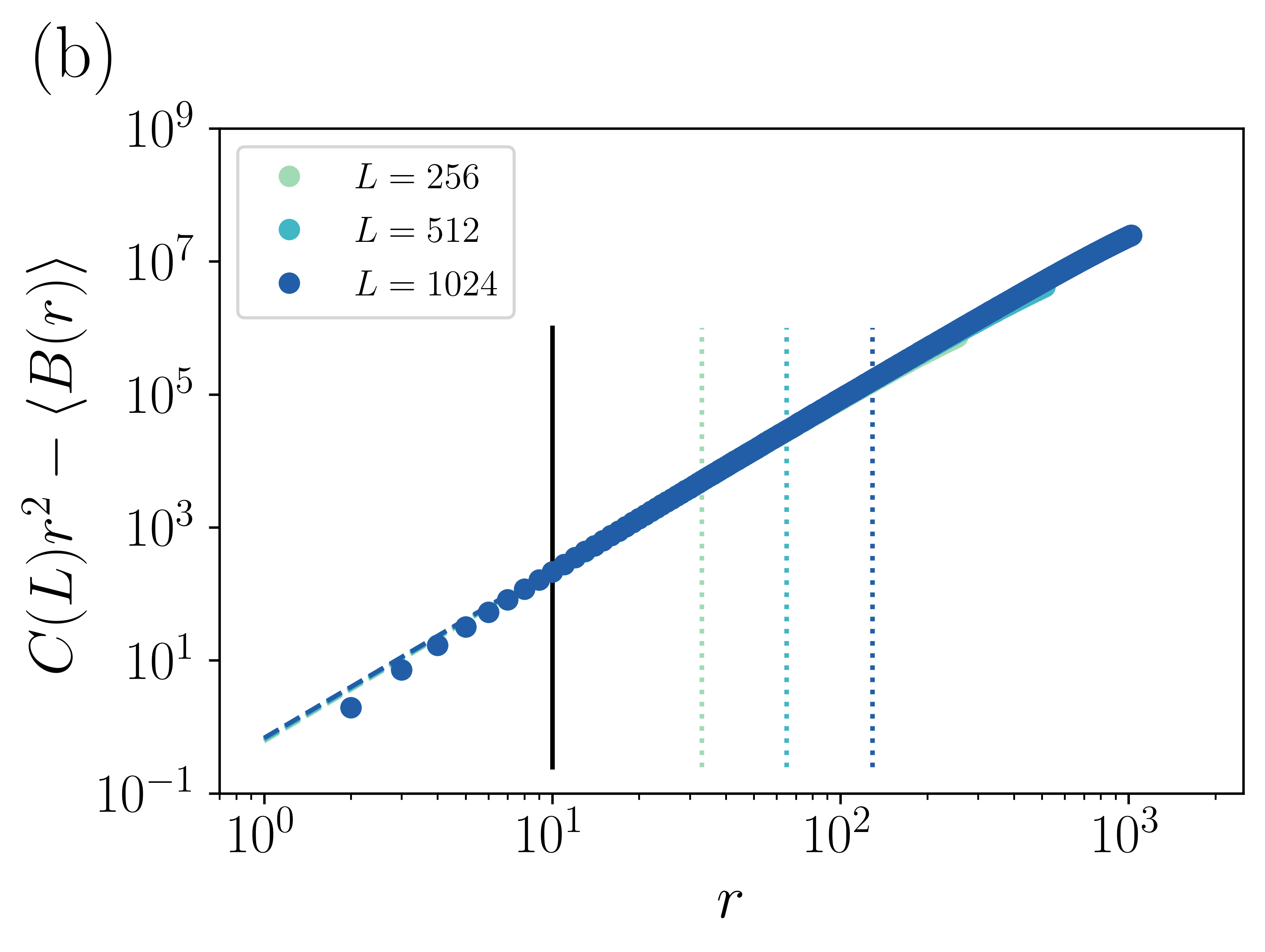}
\includegraphics[width=0.45\columnwidth]{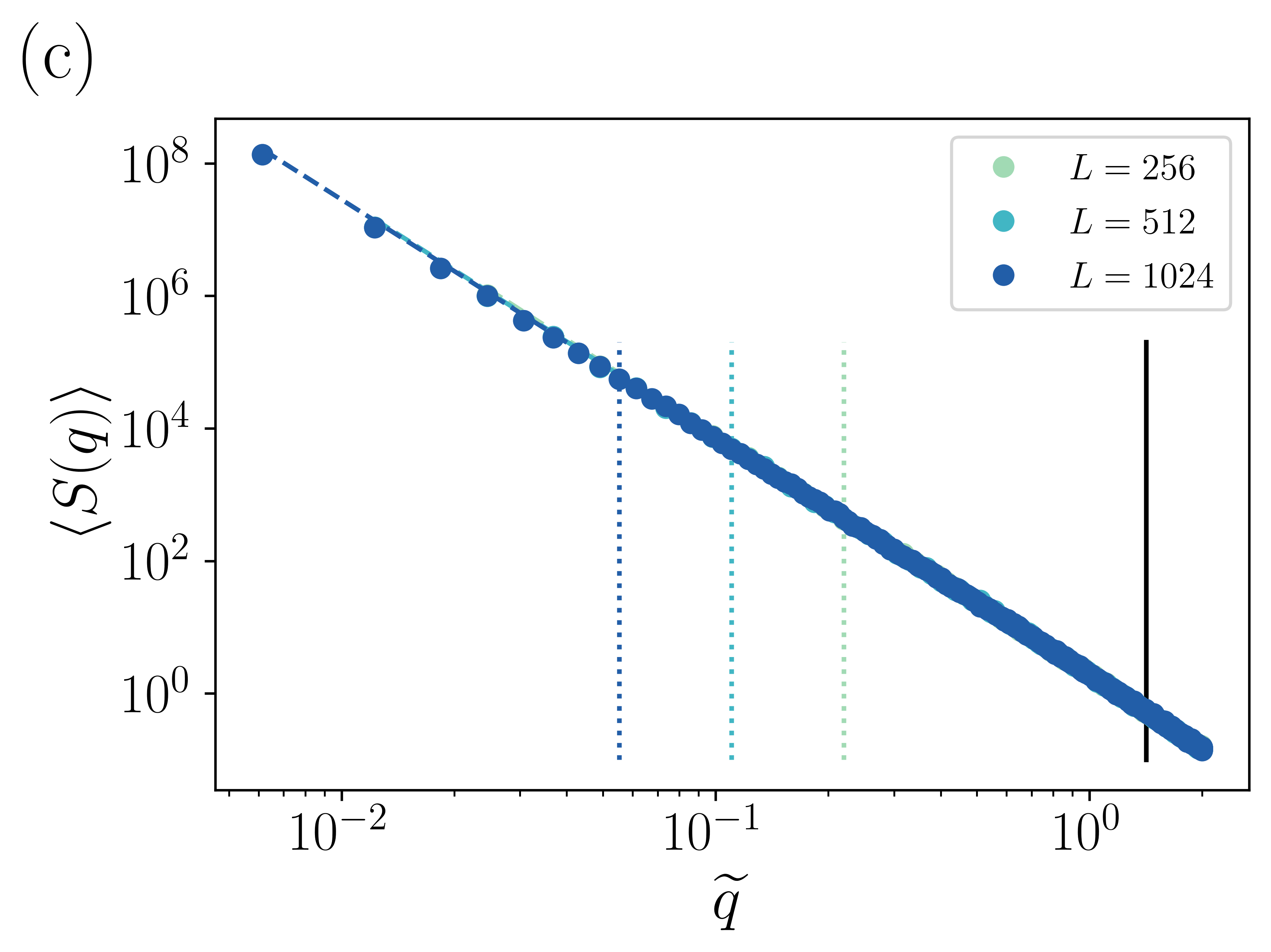}
\end{center}
\caption{
(a) Average displacement-displacement autocorrelation function ${\langle B(r) \rangle}$, (b)~scaling correction due to the super-roughness behavior, and (c) average structure factor ${\langle S(q) \rangle}$ for the driven elastic line at depinning.
The solid vertical lines indicate the common small length scale bound of the fitting ranges, while the large length scale bounds depend on the system size and are indicated by the vertical dotted lines.
}
\label{fig_corr_dep}
\end{figure}

Since the expected value for the roughness exponent ${\zeta_{\mathrm{dep}}=1.25}$ is larger than 1, interfaces are super-rough and local real-space correlation functions are not expected to recover the correct value. Indeed, in this case the local roughness exponent gives $\zeta = 1$~\cite{lopez_pre_97_anomalous_scaling, lopez_pre_98_anomalous_scaling}, as discussed in Sec.~\ref{sec_br_superrough}. Figure~\ref{fig_corr_dep}(a) shows the roughness function, where a small but noticeable vertical shift is observed when the system size changes. This is a signature of anomalous scaling corresponding to $\zeta_{\mathrm{local}} \neq \zeta_{\mathrm{global}}$~\cite{lopez_pre_97_anomalous_scaling, lopez_pre_98_anomalous_scaling}.
As discussed in Sec.~\ref{sec_br_superrough} for super-rough interfaces with an exponent ${\zeta>1}$, there is an extra independent parameter in the mathematical expression of $B(r)$ given by \eqref{equ_br_superrough}, $C(L)$ and the determination of the value of the roughness parameters cannot therefore be performed by a simple lest-square-fitting procedure.
However, this difficulty can in practice be overcome by estimating the ${C(L)}$ constant by extrapolating ${B(r)/r^2}$ to small length scales $r$ (see \eqref{equ_br_superrough}).
The scaling of ${\langle B(r) \rangle}$ is shown in \figref{fig_corr_dep}(b) for ${L=256}$, $512$, and $1024$.
The size-dependent parameter $C(L)$ is obtained using the first point of ${\langle B(r) \rangle}$ as an approximation, \textit{i.e.} $C(L) \approx \langle B(\delta z) \rangle/\delta z^2$ with $\delta z =1$. Figure~\ref{fig_corr_dep}(b) presents $C(L)-\langle B(r) \rangle/r^2$ against $r$. As observed, data presents power-law behavior independent of the system size in a finite range, scaling as $\sim r^\zeta$ according to \eqref{equ_br_superrough}. The fitted roughness exponents are reported in Table~\ref{tab_dep} and are a bit larger than the expected value ${\zeta_{\mathrm{dep}}=1.25}$. The value of $A_0$, as a measure of roughness amplitude, is also reported in the table.

The structure factor is shown in \figref{fig_corr_dep}(c), where it can be observed that a power-law behavior correctly describes the numerical data, without need of corrections due to anomalous scaling, as expected. From the fitting, the values of the roughness parameters $\zeta$ and $S_0$ are obtained and presented in Table~\ref{tab_dep}. The value of the obtained roughness exponent $\zeta$ are slightly above of the expected value, ${\zeta_{\mathrm{dep}}=1.25}$, and compares well with the one obtained using the roughness function (\figref{fig_corr_dep}(b)).
When comparing with Ref.~\cite{Ferrero2013}, which reported ${\zeta=1.25}$ using the structure factor for a system with $L = 2048$, we notice that we have used the discretization-corrected wave vector $\widetilde{q}$ (see Sec.~\ref{sec_defs}) and that we have fixed a fitting range to be consistent with the analysis presented in Secs.~\ref{sec_num_th} and \ref{sec_num_equ}, while in Ref.~\cite{Ferrero2013} the fit was done for small $q$ values directly.

\begin{figure}
\begin{center}
\includegraphics[width=0.48\columnwidth]{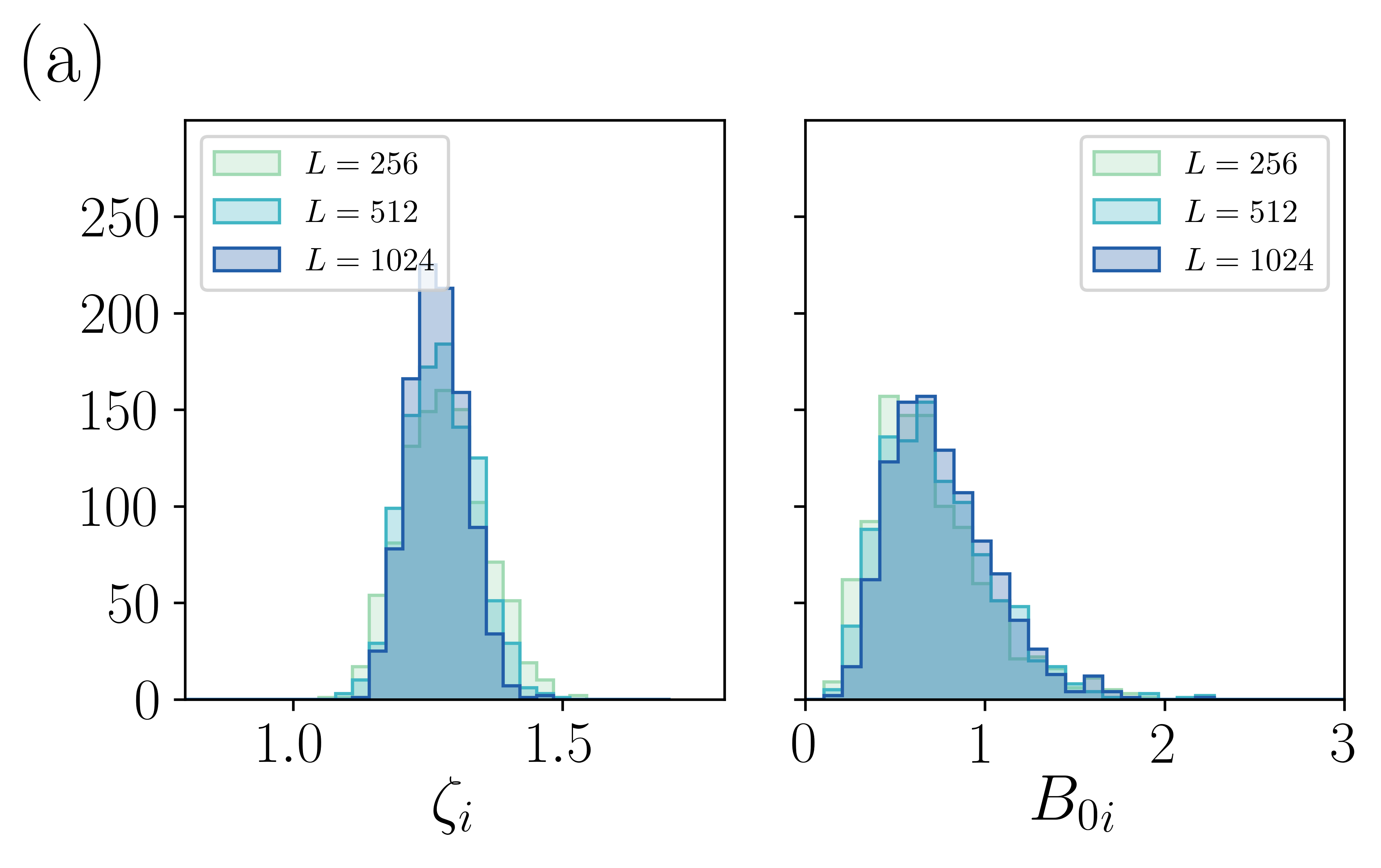}
\includegraphics[width=0.48\columnwidth]{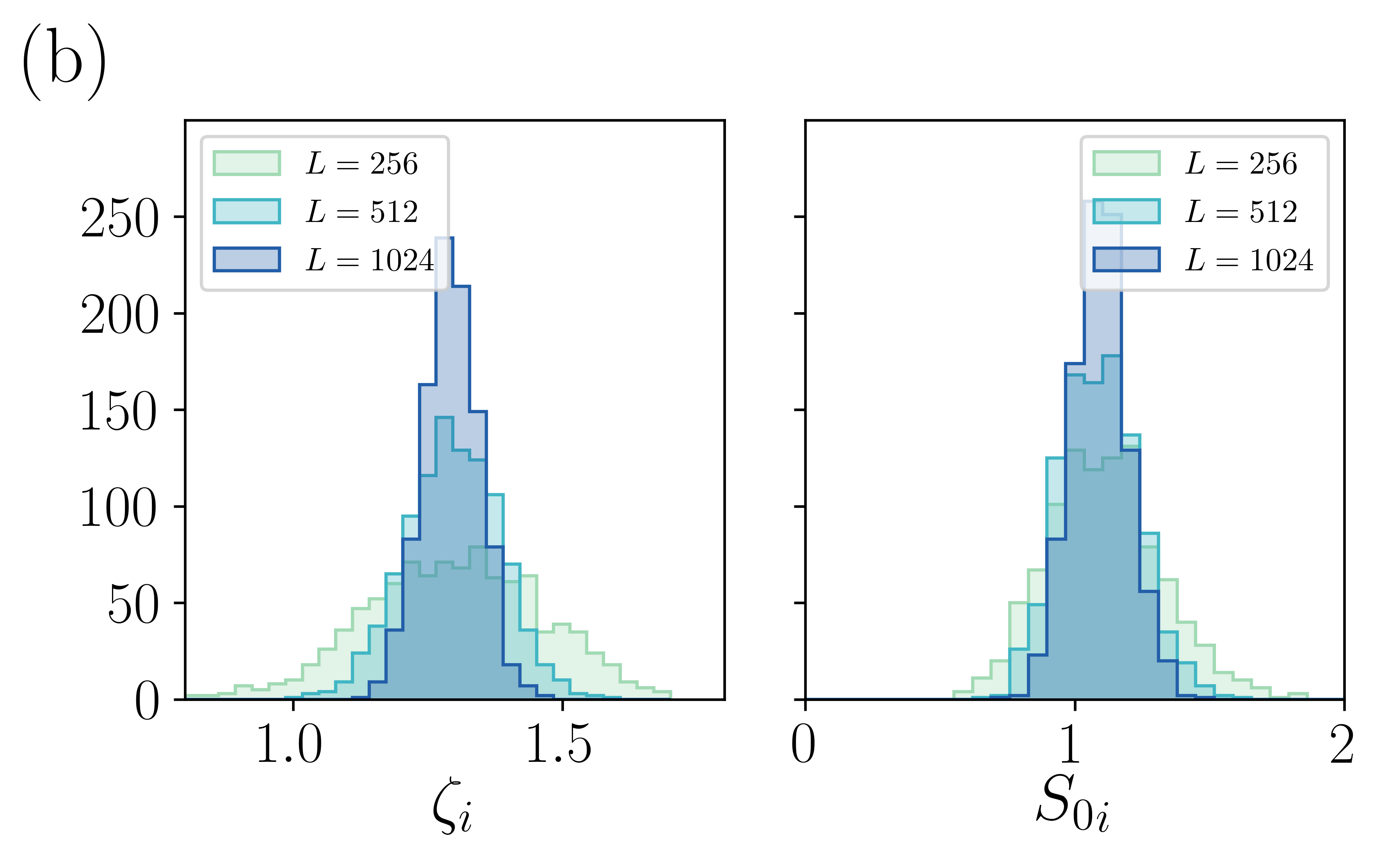}
\end{center}
\caption{
Histograms of the roughness exponent obtained using (a)~${B(r)}$ and (b)~${S(q)}$ for individual numerical interfaces corresponding to driven elastic lines at depinning.
The mean value and the variance of each distribution are indicated in Table~\ref{tab_dep}. 
}
\label{fig_dist_dep}
\end{figure}

The same fitting procedure, appropriate for super-rough interfaces, can be applied to individual realization of the depinning interfaces. Using the roughness function and considering its anomalous behavior individual values for $\zeta_i$ and ${A_0}_i$ are obtained, whose distributions are presented in \figref{fig_dist_dep}(a).
Fitting individual structure factors, values for the roughness parameters, $\zeta_i$ and ${S_0}_i$, are obtained and their distributions are presented in \figref{fig_dist_dep}(b).
The mean values for the roughness exponents, coming from both the roughness function and the structure factor, are reported in Table~\ref{tab_dep} and are slightly larger than the expected value, ${\zeta_{\mathrm{dep}}=1.25}$, considering the error bars coming from the distribution of the data.

\begin{table}[!th]
\begin{center}
\begin{tabular}{|c|c|c|c|c|}
\multicolumn{5}{c}{Driven elastic line at critical depinning (${\zeta_{\mathrm{dep}}=1.25}$}) \\
\hline
\multicolumn{2}{|c|}{$L$} & 256 & 512 & 1024 \\
\hline\hline
$\langle B(r) \rangle$ & $\zeta$ & $1.289 \pm 0.001$ & $1.281 \pm 0.001$ & $1.2747 \pm 0.0007$ \\
& $A_0$ & $0.607 \pm 0.004$ & $0.643 \pm 0.004$ & $0.681 \pm 0.003$ \\
\hline
$\langle S(q) \rangle$ & $\zeta$ & $1.309 \pm 0.007$ & $1.297 \pm 0.004$ & $1.292 \pm 0.002$ \\
& $S_0$ & $1.93 \pm 0.01$ & $1.931 \pm 0.009$ & $1.937 \pm 0.006$ \\
\hline
$B_i(r)$ & $\overline{\zeta}$ & $1.280 \pm 0.002$ & $1.274 \pm 0.002$ & $1.269 \pm 0.002$ \\
& $\overline{A_0}$ & $0.70 \pm 0.01$ & $0.73 \pm 0.01$ & $0.760 \pm 0.009$ \\
\hline
$S_i(q)$ & $\overline{\zeta}$ & $1.300 \pm 0.005$ & $1.297 \pm 0.003$ & $1.293 \pm 0.002$ \\
& $\overline{S_0}$ & $1.109 \pm 0.007$ & $1.092 \pm 0.005$ & $1.092 \pm 0.003$ \\
\hline
\end{tabular}
\end{center}
\caption{
Roughness parameters for the driven elastic line at critical depinning. Reported values for $\zeta$, $A_0$ and $S_0$ are obtained by fitting power-law behavior to average roughness function and structure factor, $\langle B(r) \rangle$ and $\langle S(q) \rangle$, respectively. $\overline{\zeta}$, $\overline{A_0}$ and $\overline{S_0}$ are the mean values of the distributions of individual values for each interface.
We considered ${N=1000}$ independent disorder realizations.
\label{tab_dep}
}
\end{table}

\section{Discussion and conclusion}
\label{sec_concl}

\begin{figure}
\begin{center}
\includegraphics[width=0.48\columnwidth]{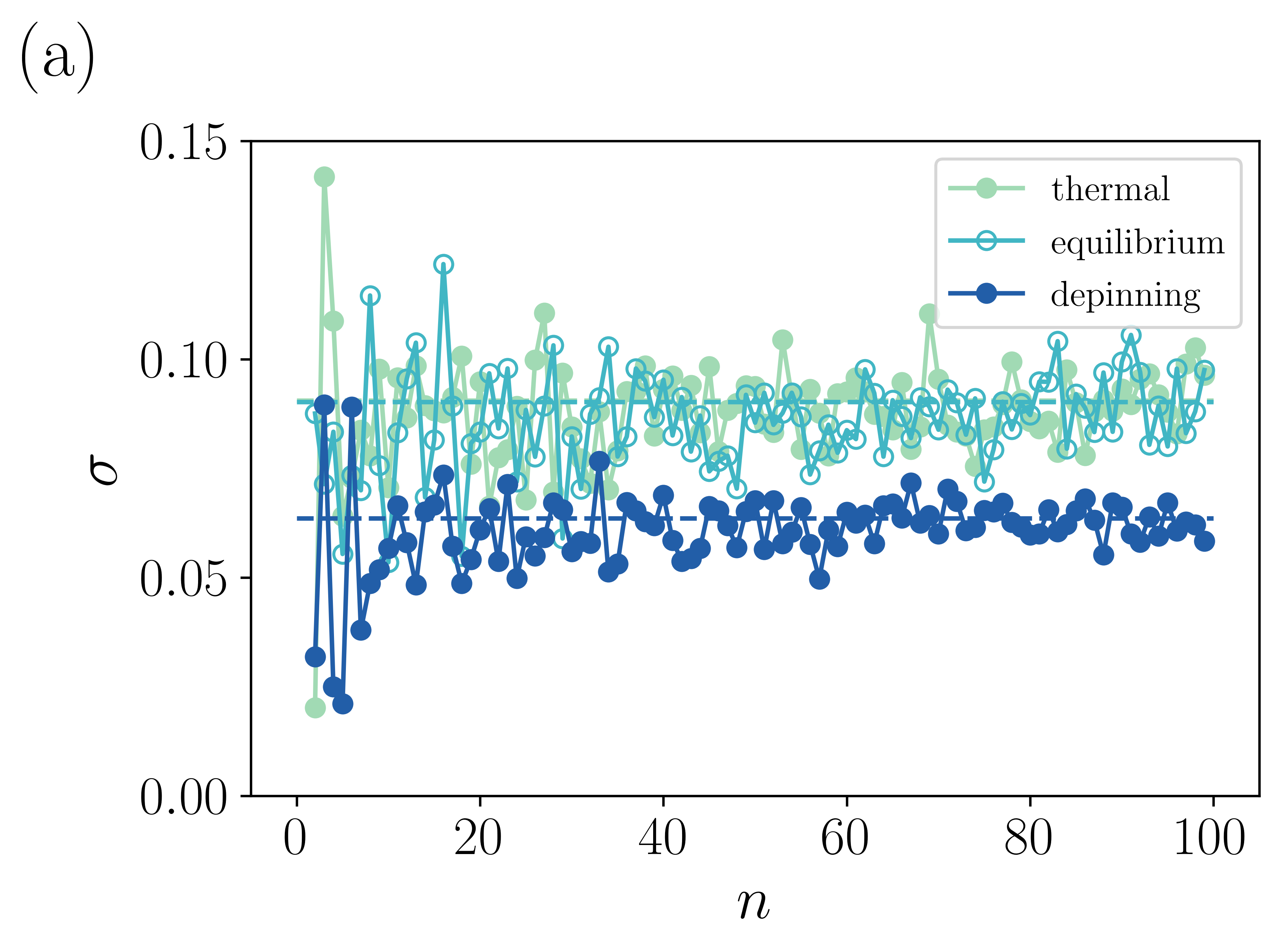}
\includegraphics[width=0.48\columnwidth]{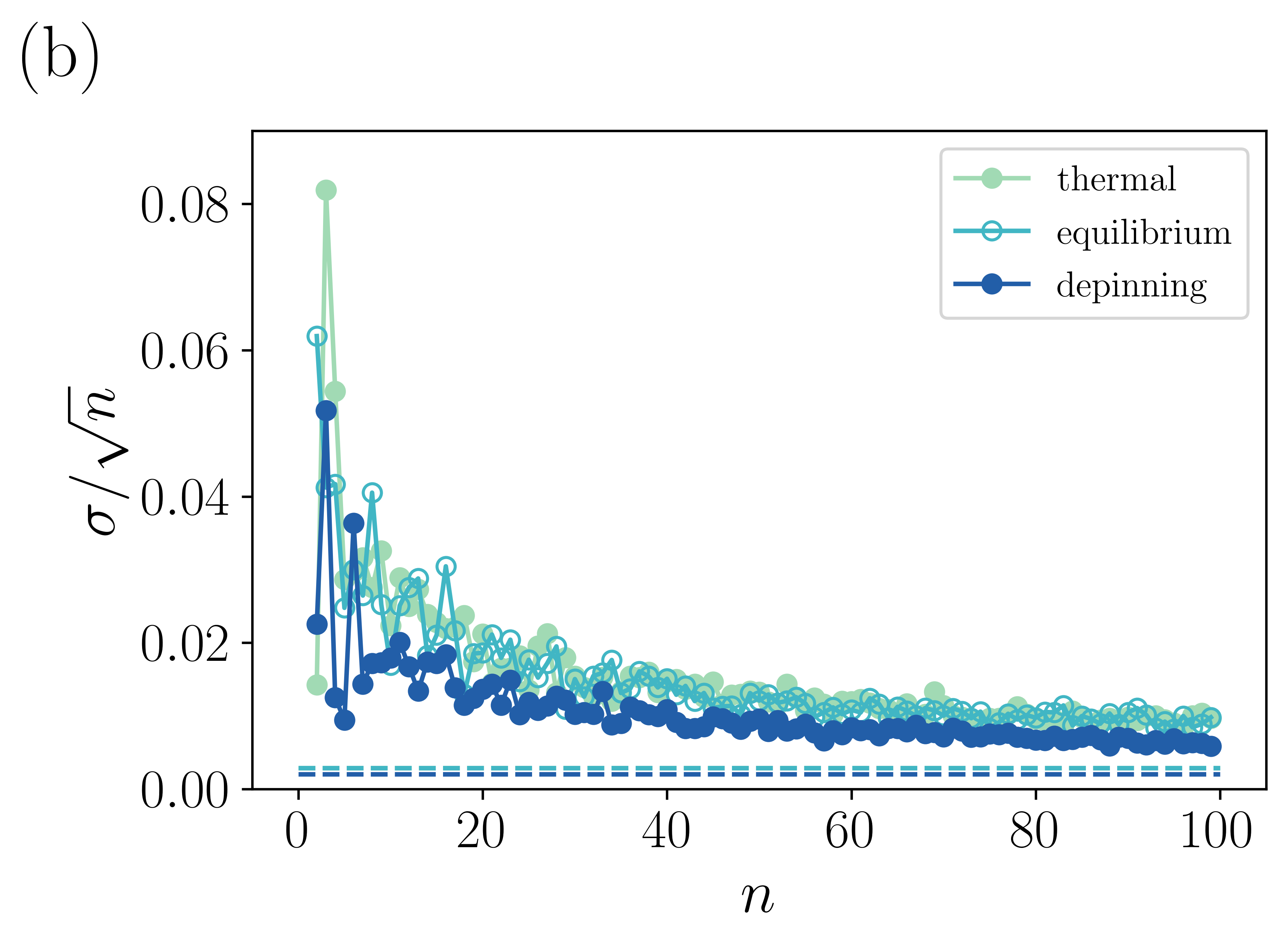}
\end{center}
\caption{ Evolution of the dispersion of the data with the size of the sample for a system size ${L = 500}$ and for the three different cases studied here (thermal, equilibrium and depinning).
(a)~Standard deviation of the distribution of roughness exponent as a measure of its width, to be compared to ${\zeta\in\lbrace 1/2, 2/3, 1.25\rbrace}$ respectively.
(b)~Error of the mean of the distribution of the roughness exponent.
Dashed lines correspond to the value for ${N = 1000}$, the largest sample studied here.
}
\label{fig_sigma}
\end{figure}

The main result presented here is the fact that the distribution of roughness parameters are considerably broad. Using model systems in three different universality classes this revealed to be a general feature of fluctuating interfaces. Since distributions are rather wide, values of roughness parameters for independent interfaces may differ considerably beyond the error bar of their mean.
In particular, the roughness exponent is broadly distributed and then the differentiation between universality classes should not be based on single finite-size interface profiles. For example, a single experimental observation of a roughness exponent $\zeta = 0.58$ would not be enough to discern between thermal Edwards-Wilkinson or equilibrium quenched Edwards-Wilkinson universality classes since the typical width of the distributions is of the order of the difference between exponents.
Furthermore, the width of the distribution of roughness exponents depends on the size of the sample, \textit{i.e.}~the number of independent individual realizations. For example, lets consider a fraction of the whole set of $N$ independent configurations composed of $n<N$ independent realizations. Figure~\ref{fig_sigma}(a) shows, as a measure of the width of the distribution, the standard deviation $\sigma$ as a function of the sample size $n$, corresponding to the distribution of $\zeta_i$ for system size $L = 512$ and for the three model systems studied here. Each sample of $n$ interface profiles were randomly taken from the $N=1000$ configurations. As observed in the figure, the width of the distribution strongly fluctuates for small $n$ and then converges after using tens of configurations (dashed lines correspond to ${\sigma(n=N)}$). Since $\sigma$ tends to a constant value, of order $0.1$, the error of the mean of the distribution, $\sigma/\sqrt{n}$, decreases as $1/\sqrt{n}$, as shown in Fig.~\ref{fig_sigma}(b). Notice that for a $10 < n < 40$ the value of the error of the mean is in the range $(0.01,0.02)$, giving thus a reasonable bound for experimental measurements of the roughness exponent.

In addition, our results convincingly show that the real-space displacements autocorrelation function can be used to determine the global roughness exponent of super-rough interfaces, via the anomalous scaling of $B(r)$ [\eqref{equ_br_superrough}].
As demonstrated for surface fractures, super-roughening has a significant impact on the morphology of the interface~\cite{lopez_pre_98_anomalous_scaling}.
Since the roughness function ${B(r)}$ is the easiest quantity to compute in experiments, \eqref{equ_br_superrough} provides a convenient way to assess the possibility of super-roughening  without having to compute the global width or structure factor.

In summary, we have shown numerical evidence for the broad distribution of individual values of roughness parameters. For this purpose, we have used both real-space and reciprocal-space correlation functions and three model systems belonging to different universality classes. This information should be taken into account when experimentally measuring roughness parameters, particularly when finite size effects are important.
Our results reveal an important but often overlooked property of roughness characterization: The measured roughness exponent originates in wide size-dependent distributions. This should always be taken into account when evaluating the roughness exponent for a given problem. For instance, when reporting the roughness exponent, a number of the order of tens of independent realizations of domain walls should be considered to guarantee statistical convergence to a meaningful average value. This result should prompt a reevaluation and development of detailed experimental protocols to assure statistical independence and finite sample size of domain wall configurations.
Such protocols would be particularly relevant for ferromagnetic and ferroelectric domain walls, since these experimental interfaces usually combine the issues of finite resolution, finite size, and limited number of experimental interfaces. However, our results are more broadly of interest for any experimental or numerical interfaces that could be described within the frame of disordered elastic systems.

\ack
The authors acknowledge enlightening discussions with J.~Curiale, T.~Giamarchi, P. Guruciaga, M.~Granada and A.B.~Kolton.
JG and PP gratefully acknowledge financial support from the Fondation Ernst et Lucie Schmidheiny and SNSF grant 200021\_153174. EA acknowledges support from the Swiss National Science Foundation under the SNSF Ambizione Grant PZ00P2\_173962. SB acknowledges support from the Grant No. PICT 2017-0906 from the Agencia Nacional de Promoción Científica y Tecnológica, Argentina.
\\

\bibliography{refs_strings,biblioEwok}
\bibliographystyle{iopart-num.bst}

\end{document}